\documentclass[10pt,letterpaper]{article}
\usepackage[]{geometry}

\usepackage{amsmath,amssymb}
\usepackage{mathtools}
\usepackage{changepage}

\usepackage{textcomp,marvosym}

\usepackage{cite}
\usepackage[hyphens]{url} 

\usepackage{nameref,hyperref}

\usepackage[right]{lineno}
\usepackage{tabularx}

\usepackage[table]{xcolor}

\usepackage{array}

\usepackage{longtable}

\usepackage{amstext}
\usepackage{pgfplotstable}
\pgfplotstableset{
  begin table=\begin{longtable},
  end table=\end{longtable},
}

\pgfplotsset{compat=newest}
\usepackage{booktabs}
\usepackage{amsmath,amssymb,amsbsy}
\usepackage{graphicx}
\usepackage{epsfig}
\graphicspath{{FIG/}}
\usepackage{wrapfig}
\usepackage[final]{changes}

\reversemarginpar

\usepackage{lastpage,fancyhdr,graphicx}

\graphicspath{{FIG/}}

\begin{document}


\begin{flushleft}
{\Large
\textbf\newline{
%
%
The role of APOBEC3-induced mutations in the differential evolution 
of monkeypox virus
} }


Xiangting Li\textsuperscript{1\Yinyang},
Sara Habibipour\textsuperscript{2\Yinyang},
Tom Chou\textsuperscript{1,3\textcurrency},
Otto O. Yang\textsuperscript{2},
\\
\bigskip
\textbf{1} Department of Computational Medicine, UCLA, Los
Angeles, CA, United States
\\
\textbf{2} Depts. of Medicine and Microbiology,
Immunology, and Molecular Genetics, UCLA, Los Angeles, CA, United States
\\
\textbf{3} Department of Mathematics, UCLA, Los
Angeles, CA, United States
\\
\bigskip

%
%
\Yinyang These authors contributed equally to this work.






\end{flushleft}
\section*{Abstract}
Recent studies show that newly sampled monkeypox virus (MPXV) genomes
exhibit mutations consistent with Apolipoprotein B mRNA Editing
Catalytic Polypeptide-like3 (APOBEC3)-mediated editing, compared to
MPXV genomes collected earlier. It is unclear whether these single
nucleotide polymorphisms (SNPs) result from APOBEC3-induced editing or
are a consequence of genetic drift within one or more MPXV animal
reservoirs. We develop a simple method based on a generalization of
the General-Time-Reversible (GTR) model to show that the observed SNPs
are likely the result of APOBEC3-induced editing. The statistical
features allow us to extract lineage information and estimate
evolutionary events.
%


\section*{Introduction}

Monkeypox (MPX), or Mpox, is a viral zoonotic disease that
has been reported sporadically in western and central Africa. It is
caused by the orthopoxvirus MPXV, which is related to Variola Virus,
the past cause of the smallpox pandemic. MPX resides in a yet
unidentified animal reservoir, and whether it has achieved endemicity
in humans is unclear \cite{who:online,eucdc:online,uscdc:online}.

In 1970, the first human case of monkeypox was reported in the
Democratic Republic of the Congo (DRC) \cite{Breman1980human}. Since
then, it has been suggested that MPX may have become endemic in
the DRC, and spread to several other Central and West African
countries \cite{Bunge2022feb}. Outside of Africa, there were sporadic
cases of MPX reported from 2003 to 2021, until the recent 2022
worldwide outbreak \cite{who:online,eucdc:online,uscdc:online}.
These previous sporadic cases were shown to be
related to travel to Africa \cite{Bunge2022feb}. While cases in the
most recent outbreak did not involve travel to or from Africa, the
sequences from the 2022 outbreak were found to be clustered with
2018-2019 cases \cite{Isidro2022jun}. Moreover, O'Toole and Rambaut
first reported signs of Apolipoprotein B mRNA Editing Catalytic
Polypeptide-like3 (APOBEC3)-mediated editing in those sequences
\cite{Rambaut2022a,Rambaut2022b,Rambaut2022c}. 

APOBEC3 is a group of human enzymes of the innate immune system
capable of inhibiting certain types of viruses through deaminating
cytosine to uracil, causing a G-to-A mutation in the complement strand
when it is synthesized \cite{Sadeghpour2021jul}.  Most human APOBEC3
molecules tend to deaminate TC dinucleotides, except APOBEC3G, which
prefers CC dinucleotides
\cite{Yu2004apr,Stenglein2010jan,Hultquist2011nov,Burns2013feb,Beale2004mar}.
Compared to MPXV sequences collected in the 1970s, the single
nucleotide polymorphisms (SNPs) in the 2022 outbreak were found to be
biased for the pattern ``TC $\to$ TT'' and ``GA $\to$ AA''
\cite{Rambaut2022a,Rambaut2022b,Rambaut2022c}. O'Toole and Rambaut
hypothesized that the observed SNPs in the 2022 outbreak were the
result of APOBEC3-induced editing, and that they are evidence of
within-human evolution of MPXV
\cite{Rambaut2022a,Rambaut2022b,Rambaut2022c}.

Subsequent phylogenetic analysis of the genomes collected between 2017
and 2022 revealed significantly \replaced{high}{higher} frequencies of
APOBEC3-associated SNPs, further supporting the hypothesis of
within-human evolution of MPXV
\cite{Isidro2022jun,gigante2022multiple}.  In addition, all the 2022
sequences appeared to cluster together and little evidence of
APOBEC3-related mutations was found prior to 2016. In the context of
those studies, the APOBEC3-associated SNPs were defined as specific
``dinucleotide mutation'' of the form ``TC $\to$ TT'' and ``GA $\to$
AA.''

Because DNA evolution is a complex process involving mutation and
selection, \added{there may be a number of different explanations for
  the higher fraction of APOBEC3-associated SNPs observed in later
  sequences; we wish to quantify the influence of APOBEC3 editing by
  factoring out the effects of these other mechanisms.}  If the
average mutation rates \textit{per site} of the TC $\to$ TT type are
intrinsically higher than other types of mutations, independent of
APOBEC3 activity, then a high fraction of APOBEC3-associated SNPs
could be explained without invoking APOBEC3 involvement. Here, we take
``intrinsically'' to mean APOBEC3-like mutations
\cite{You2000nov,Hussein2005mar,Budden2013jan} caused by mutagens
unrelated to APOBEC3 such as UV radiation that induce ``C $\to$ T''
mutations \cite{Drouin1997nov}.
%
%
%
%
\replaced{Besides the mutagens, the observed mutations themselves may
  confer higher fitness to MPXV, for instance, by enabling the virus
  to evade immune surveillance and attack cells more
  effectively}{These mutations could increase MPXV fitness by
  promoting immune evasion and cellular invasion,} \added{and by
  exploiting tRNA abundance via codon usage
  biases\cite{Parvathy2021nov}. Mutations that are indistinguishable
  from those induced by APOBEC3 activity can thus be selected even in
  the absence of APOBEC3.}

\deleted{The prevalent APOBEC3-associated
SNPs may also stem from numerous nonlethal mutable APOBEC3-related
sites, where the mutation rate \textit{per site} mirrors the average
rate at other sites.}
\deleted{In addition to these previously discussed factors,
  such as other mutagens and selection, APOBEC3 family proteins also
  exist in primates other than humans. Although presumably due to
  fierce evolutionary competition between virus and host, APOBEC3
  genes have rapidly evolved within the primate lineage. A recent
  study showed that APOBEC3B, APOBEC3D, and APOBEC3F are the youngest
  members in human and probably originated in the common ancestor of
  Catarrhini.  Therefore, we should also account for possible
  contributions from APOBEC3 family proteins in the unknown animal
  reservoir of MPXV.}


\added{To distinguish APOBEC3-induced mutations from
  other high-frequency mutations, we can establish a baseline
  by investigating early (pre-2016) MPXV genomes.  However,
  time-varying mutation rates, uncertainty in sampling time points,
  and a limited number of early sequences complicates ancestor
  sequence reconstruction and mutation event timing.  To circumvent
  these uncertainties, we develop a method to quantify
  \textit{relative} mutation rates and find (i) that even in genomes
  collected before 2016, the TC $\to$ TT mutation rate relative to
  other types is higher than average and (ii) that TC $\to$ TT
  mutations are even more abundant in genomes collected after
  2016. Mutagens induced by UV radiation may account for the higher
  mutation rate of TC $\to$ TT in the pre-2016 genomes, while our
  hypothesis is that the acceleration of TC $\to$ TT mutations in the
  later (post-2016) genomes arise from human-specific APOBEC3
  editing.}

To factor out selection, we examine the evolution of synonymous SNPs
which do not alter the amino acid sequence and have minimal influence
on the viral fitness.  \added{Even within different synonymous
  mutations, codon usage bias may arise; however, a previous study on
  the codon usage bias of monkeypox virus collected before 2016
  concluded that the genome-wide frequency of nucleotide at the third
  codon position was not dependent on nucleotides at the first or
  second position~\cite{Karumathil2018jan}, suggesting that natural
  selection did not favor particular codons in the MPXV genomes. We
  count the number of synonymous mutations and synonymous
  APOBEC3-relevant mutations by first identifying all mutations at all
  sites, then identifying specific mutations that are synonymous.}
\added{We also consider the preceding nucleotide, which may or may not
  be within the same codon, to count the number of APOBEC-relevant
  dinucleotide mutations such as TC $\to$ TT.}


Our model allows for variation in the ``raw'' or absolute mutation
rate across different genomic sites, lineages, and generations. For
example, one specific lineage of the virus may acquire a beneficial
mutation at a specific time, allowing the virus to replicate faster
and thus increasing the subsequent mutation rates in this
faster-growing lineage. However, to quantitatively understand the
evolution of synonymous SNPs, we make a key assumption that the
\textit{ratio} of rates of any two types of mutation is \replaced{the
  same across all sites, lineages, and across time, even though the
  raw rates may vary across sites, lineages and time.}{constant}
\added{Our method is designed to capture common patterns of
  \textit{relative} mutation rates shared by all members of a lineage
  sharing a common ancestor.}

The assumption of constant mutation rate ratios underpins all Markov
models of DNA nucleotide substitutions, such as the JC69
\cite{JUKES1969} and GTR \cite{Tavare1986} models. Evolution models
used in maximum likelihood and Bayesian inference phylogenetic methods
also implicitly rely on the constant rate ratio assumption
\cite{Felsenstein1981nov,Nguyen2014nov,Felsenstein2001oct} \added{as
  typical phylogenetic methods are derived based on relative rates of
  mutations and do not involve absolute time.}  Thus, our key
assumption of constant relative mutation rates is commonly used.
Under such an assumption, the number of the observed mutations of a
given type will be proportional to the number of another type in the
same way across genomes from different lineages and generations. The
assumption of linear proportionality is similar to that used in a
strict molecular clock theory, which predicts that the mean number of
(neutral) mutations increases linearly with physical time.
\replaced{This similarity inspires us to integrate phylogenetic models
  with principles derived from molecular clocks. We aim to correlate
  the number of synonymous mutations with a conceptualized relative
  time scale. We refer to the linear proportionality of the number of
  a specific mutation to the total number of synonymous mutations as
  the ``relative molecular clock''.}{This strict clock assumption is
  too strong for practical use. Relaxed clock models have been
  formulated to allow the evolutionary rate to vary across lineages.
  In these relaxed clock models, the linearity between the number of
  mutations and time is lost. By contrast, our approach is more
  flexible than either the strict or relaxed molecular clock
  models. By comparing mutations of different types, we
  ``recalibrate'' the relaxed molecular clock to the number of
  mutations of another type, instead of evolutionary time or number of
  generations. This restores the linearity of the clocks.
  Visualization of different genomes in the same coordinate system and
  validation of the linearity provide a direct and intuitive way to
  validate our key assumption. We will refer to our method as the
  ``recalibrated molecular clock''.}

\deleted{Inferring evolutionary information from the
  ratio of mutations of different types is not a new idea. For
  example, the ratio of nonsynonymous mutations to synonymous
  mutations $\mathrm{d}_N/\mathrm{d}_S$ is often used to infer the
  selective pressure on the protein-coding regions. Methods to
  identify site-specific mutation rate-shifts have also been
  proposed. Application of these methods generally rely on an
  evolution model to reconstruct the phylogenetic trees via maximum
  likelihood or Bayesian inference. This requires considerable
  computational resources and does not provide an intuitive way to
  understand how well the model fits the data. By contrast, our method
  does not rely on the phylogenetic trees, requires only sequence
  alignments against a common reference sequence, and provides an
  intuitive goodness-of-fit measure.}

In the next (Materials and Methods) section, we formalize \added{and
  validate} our ideas \added{by first introducing a complete DNA
  substitution model with raw rates and absolute time.  The raw rates
  $v$ will cancel each other when we consider ratios of mutations in
  our final results. We also introduce an additional assumption of
  independence between sites which allows us to easily model the
  distribution of mutations using a simple binomial distribution and
  perform statistical hypothesis testing based on this
  distribution. In particular, our method does not require any
  knowledge of the ancestor sequence.}

Using representative MPXV genomes collected from the National Center
for Biotechnology Information (NCBI), we demonstrate that our
\replaced{relative}{recalibrated} molecular clocks can
describe the distribution of the fraction of APOBEC3-associated SNPs
to the total number of synonymous SNPs. The genomes collected before
and after 2016 clearly separate into two groups, with each group
exhibiting a shared relative rate of APOBEC3-associated mutations. We
surmise that these two groups of genomes were derived from two
different evolutionary environments. The genomes collected before 2016
were likely direct descendants from the animal reservoir, while the
genomes collected after 2016 were likely derived from the human host.
\added{Given that MPXV jumped from the animal reservoir to
humans~\cite{Huang2022nov}, we can infer statistical properties such
as the number of synonymous mutations between the common ancestor of
the post-2016 genomes (which has not yet undergone APOBEC3-editing)
and a reference genome.}
%
%
This statistically inferred common ancestor can be older than the last
common ancestor of the same group and provide a better estimate of the
time of zoonotic MPXV transmission. Based on the inferred relative
rates and additional independence assumption, we also developed a
statistical test to determine whether a given genome can be considered
as the common ancestor of the post-2016 genomes.
%
%
Evolution of other viruses with APOBEC3-induced editing can also be
analyzed using the models presented here.

\section*{Materials and Methods}

\subsection*{DNA evolution model}
\replaced{We construct a relative clock by using the number of
  synonymous SNPs.}{calibration of a relaxed molecular clock}
\replaced{To relate our method to both molecular clock and
  phylogenetic models, we start with a formal substitution model with
  raw mutation rates and absolute time.} For the sake of simplicity,
we first consider a variant of the GTR model \cite{Tavare1986} that
accounts for local dinucleotide context and mutation rates that are
homogeneous across the genome, all lineages, and all generations. We
then discuss how relaxing these homogeneity assumptions, by proper
rescaling of absolute time, can still lead to a
constant-\textit{relative}-mutation-rate structure.

\added{Single-site DNA evolution models typically use a 4$\times$4
  matrix whose elements describe substitutions among the four
  nucleotides. For example, the entry $\mathbf{V}_{\rm AC}$ might
  describe the A $\to$ C mutation rate, defined by the expected number
  of mutations per site per unit time. The rate matrix ${\bf V}$ can
  be decomposed into the product of the mutation probabilities per
  site per replication cycle and the raw replication rate.  The JC69
  model assumes each available mutation is equally likely to occur and
  uses a single parameter $v$ that describes the overall mutation
  rate.  Therefore, in JC69, $\mathbf{V}= v \mathbf{Q}$ where
  $\mathbf{Q}$ is a normalized substitution matrix $\mathbf{Q}_{XY}=
  \frac{1}{3}$ if $X\neq Y$ and $\mathbf{Q}_{XX}=-1$.  Similarly, the
  rate matrix $\mathbf{V}$ in the GTR model can be decomposed into
  $\mathbf{V} = v \mathbf{Q}$, where $v$ and $\mathbf{Q}$ are both
  parameters of the model.}

To reformulate the DNA evolution model to include local dinucleotide
context, we extend $\mathbf{Q}$ to a 16$\times$16 matrix with entries
describing the normalized rate of mutations between each pair of
dinucleotides.  In the following subsections, we allow $v$ to be
dependent on genomic sites $x$, lineage $(i)$, and time $t$, while
$\mathbf{Q}$ is a constant matrix to be shared by all \added{mutation
  sites, lineages, and generations.  Under this generalization, the
  mutation rate matrix for lineage $(i)$ at time $t$ and site $x$ is
  $\mathbf{V}^{(i)}(x,t) = v^{(i)}(x,t) \mathbf{Q}$. Because the
  mutation rate separates from the constant substitution matrix
  $\mathbf{Q}$, the ratios of mutation rates are independent of the
  mutation rate prefactor $v^{(i)}(x,t)$.}

Several factors can influence the substitution matrix, including the
DNA replication and proofreading mechanisms, the environmental factors
such as UV, pH, and temperature that may induce spontaneous base
substitution, and other proteins that actively edit the DNA sequence,
such as the APOBEC3 family of proteins.  As an orthopoxvirus, MPXV
carries its own replication proteins and replicates in the cytoplasm
of the host cell \cite{Peng2023jan}. Within the human host, the
environmental factors and host proteins are relatively
constant. Therefore, we conclude that the ratio of mutation rates are
mostly conserved, and our key assumption is reasonable.

Here, we focus primarily on synonymous SNPs. Historically, synonymous
SNPs can be considered as neutral mutations and are often associated
with the molecular clock \cite{Fitch1976}. However, selection pressure
related to codon usage \cite{Wallace2013mar} can differentially affect
synonymous mutations.  Our model does not require neutrality, but it
does require that mutations of the same type occur with similar rates
relative to other mutations. Statistically, different synonymous
mutations of the same pattern, \textit{e.g.}, TA $\to$ TT, are
considered as identical. This allows us to count the number of
synonymous mutations of the same type and infer the relative rates of
these mutations.  \added{Our consideration mostly
  involves DNA substitution models rather than codon substitution
  models\cite{Goldman1994sep,Yang2014may}. We identify synonymous
  dinucleotide mutations by the context along the whole sequence
  rather than only through synonymous sites.}

\begin{figure}[htbp]
  \centering
  \includegraphics[width=0.9\textwidth]{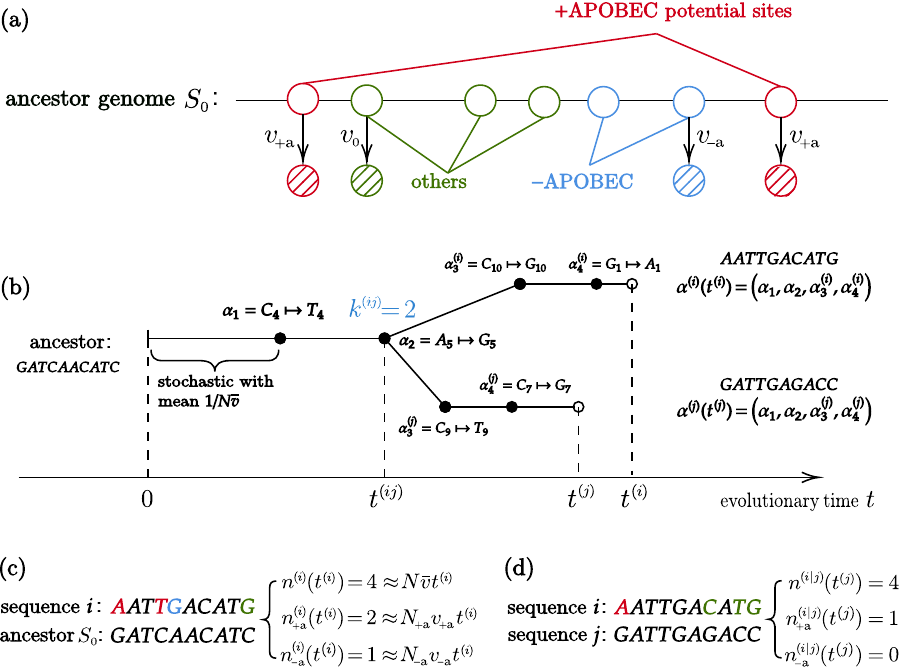}
  \caption{Schematic of the stochastic evolution model. (a) There are
    different potential synonymous mutation sites along the genome.
    Some potential mutations are APOBEC3-induced (red), some are
    reverse-APOBEC3-induced (blue), and some are neither (green).  The
    transitions (from open circles to hash-filled circles) are
    indicated by arrows. Transitions of each type occur independently
    at their type-specific transition rate. (b) An example of two
    lineages $(i)$ and $(j)$ arising from a common ancestor that shared a
    common evolutionary path until time $t^{(ij)}$. Each black dot
    represents a mutation. An open circle represents the time a
    lineage is sampled. The waiting time between two successive
    mutations are independent and randomly distributed with the same
    total rate $N \bar{v}$. The specific type of mutation is also
    randomly chosen according to the relative rates of the different
    mutation types. (c) Counts of mutations using the common ancestor
    as a reference. (d) Counting the numbers of mutations of lineage
    $(i)$ at time $t^{(i)}$ using the lineage $(j)$ at time $t^{(j)}$ as a
    reference. Since sequences share part of their evolutionary paths
    starting with the common ancestor, mutations $\alpha_1$ and
    $\alpha_2$ are not identified by using lineage $(j)$ as a
    reference. The subsequent mutations of lineage $(j)$ count towards
    $n^{(i|j)}$. For \added{completeness}, we included all mutations
    instead of only synonymous ones in the schematic sample sequences
    in (c) and (d).}
  \label{fig-schematic}
  \end{figure}

Now suppose there is a common ancestor sequence $S_0$ in which there
are a total of $N$ possible synonymous mutations,
\added{\textit{i.e.}, roughly $N/3$ codons since most
codons have 3 synonymous alternatives}, a total of $N_{\rm +a}$
possible synonymous APOBEC3-induced mutations (TC $\to$ TT and GA
$\to$ AA), and a total of $N_{\rm -a}$ possible synonymous
reverse-APOBEC3 mutations (TT $\to$ TC and AA $\to$ GA). \added{To
quantify $N$ and $N_{\rm +a}$, each possible nucleotide change that
retains the subsequent amino acid is counted as a separate synonymous
mutation.}  As shown in Fig.~\ref{fig-schematic}(a), the basic idea is
to assume that each mutation of the same type along the genome is
independent and identically distributed with a type-specific rate.

Synonymous APOBEC3-induced mutations are defined as synonymous
mutations that match the pattern TC $\to$ TT or its reverse complement
GA $\to$ AA. The reverse-APOBEC3 synonymous mutations are defined as
synonymous mutations that match the pattern TT $\to$ TC or its
complement AA $\to$ GA. Together, these two types of synonymous
mutations will be denoted ``APOBEC3-relevant'' in the following
sections. The remaining possible number of synonymous mutations $N_{0}
= N - N_{\rm +a} - N_{\rm -a}$ are non-APOBEC3-relevant.

We assume that all possible synonymous mutations of the same type
$\sigma\in\{0, {\rm +a}, {\rm -a}\}$ have the same mutation rate
$v_\sigma$ and that each mutation occurs independently of others.  For
example, the mutation rate for synonymous APOBEC3-induced mutations is
$v_{\rm +a}$, and the mutation rate for synonymous reverse-APOBEC3
mutations is $v_{\rm -a}$. The average mutation rate for synonymous
non-APOBEC3-induced mutations is $v_{0}$. The average
mutation rate across all possible synonymous mutations is thus 

\begin{equation}
    \bar{v} = \frac{N_{\rm +a} v_{\rm +a} + N_{\rm -a} v_{\rm -a} + N_{0} v_{0}}{N}.
\label{vbar}
\end{equation}

\noindent Table~\ref{table:symbols} summarizes the notation used in the
following sections.
\begin{table}[ht]
  \centering
  \small
  \begin{tabularx}{\textwidth}{X l l}
  \hline
  Name & Symbol & Typical values \added{for MPXV} \\
  \hline
  No. of possible synonymous mutations (of a given type) & $N$, ($N_{\pm\rm a}$) & 100,000 \\
  No. of observed synonymous mutations (of a given type) & $n$, ($n_{\pm \rm a}$) & 100 \\
  Mutation rate of a single nucleotide (of a given type) & $v$, ($v_{\pm\rm a}$) & -- \\
  A specific mutation out of $N$ possible mutations  & $\alpha$ & $\{1,\ldots,N\}$ \\
  No. of observed synonymous mutations up to time $t$ & $n(t)$ & 100 \\
  Sequence of mutations of $i^{\rm th}$ lineage & $\boldsymbol{\alpha}^{(i)}(t)
=(\alpha_1^{(i)},\cdots, \alpha^{(i)}_{n(t)})$ & -- \\
  $k^{\rm th}$ (chronological) mutation in lineage $(i)$  & $\alpha_k^{(i)}$ & $\{1,\ldots,N\}$ \\
  No. of observed mutations of genome $(i)$ relative to $(j)$ & $n^{(i|j)}$ & 100 \\
  Sampling time of lineage or genome $(i)$ & $t^{(i)}$ & -- \\
  Emergence time of last common ancestor of lineages $(i), (j)$ & $t^{(ij)}$ & -- \\
  No. of mutations shared by lineages $(i)$ and $(j)$ & $k^{(ij)}$ & -- \\
  \hline 
  \end{tabularx}
  \caption{Notation used in our model and analysis.}
  \label{table:symbols}
\end{table}

The different types of mutations invoked in our model
are listed in Table \ref{table:names}.

\begin{table}[ht]
\centering
\begin{tabular}{lll}
\hline Mutation type & Symbol & Synonymous Mutation Type \\ 
\hline 
Index & $\sigma$ & $\{0, {\rm +a}, {\rm -a}\}$ \\
APOBEC3-induced & $n_{\rm +a}$ & $\textrm{TC} \rightarrow
\textrm{TT}$, $\textrm{GA} \to \textrm{AA}$ \\
reverse-APOBEC3 & $n_{\rm -a}$ &$\textrm{TT} \rightarrow \textrm{TC}$,
$\textrm{AA} \to \textrm{GA}$\\
APOBEC3-relevant & $n_{\rm a} = n_{\rm +a} + n_{\rm -a}$ &
$\textrm{TC} \leftrightarrow \textrm{TT}$, $\textrm{GA}
\leftrightarrow \textrm{AA}$\\
synonymous & $n$ & all types of synonymous mutations \\
non-APOBEC3-relevant & $n - n_{\rm a}$ & all but APOBEC3-relevant
mutations \\
$\mathrm{AC} \rightarrow \mathrm{AT}$ & -- & $\mathrm{AC} \rightarrow
\mathrm{AT}$, $\textrm{GT}\to \textrm{AT}$ \\
\,\,\,\,$\textrm{A}\to {\rm C}$ & -- & $\textrm{A}\to {\rm C}$, ${\rm T}\to
  {\rm G}$\\ 
\hline
\end{tabular}
\caption{A list of the different types of synonymous mutations considered in our model.}
\label{table:names}
\end{table}


\paragraph*{Single lineage evolution.} First consider a single
lineage of MPXV genomes evolving out of the common ancestor $S_0$
starting at time $0$. In terms of the synonymous mutations, the
lineage observed at time $t$ is described by a sequence
$\boldsymbol{\alpha}(t) = (\alpha_{1},\cdots \alpha_{n(t)})$, where
each $\alpha_k \in \left\{1,..., N \right\}$ labels the location
(which also defines the mutation type) of the $k^{\rm th}$
(chronological) synonymous mutation, relative to some reference genome
(\textit{e.g.}, $S_{0}$).  Here, $n(t)$ describes the total number of
synonymous mutations that have occurred up to time $t$. An example of
two lineages $\boldsymbol{\alpha}^{(i)}$ and
$\boldsymbol{\alpha}^{(j)}$ is shown in Fig.~\ref{fig-schematic}(b).

To simplify the analysis, we also adopt an \textit{infinite sites
  assumption} \cite{Ma2008sep} in which the total number $n^{(i)}(t)$
of synonymous mutations of lineage $\boldsymbol{\alpha}^{(i)}$
satisfies $n^{(i)}(t) \ll N$. For the MPXV genomes we investigated,
the number $n$ of observed synonymous mutations is around 100, and the
total number $N$ of possible synonymous mutations is around
100,000. Therefore, we can safely assume that all observed mutations
$\alpha^{(i)}_k$ occurs on different genomic sites for different
lineages $(i)$ and chronological order $k$. Multiple mutations and
back mutations occurring on the same site are rare and neglected. We
note that the infinite sites assumption is not fundamental to our
calculation. To drop this assumption, we need to consider the relation
between the number of \textit{observed} mutations and the number of
\textit{actual} mutations. For example, if one back mutation occurs,
the observed number of mutations will be two less than the actual
number of mutations. Given the total number of actual mutations, one
can calculate the expected number of observed mutations. Inversely,
when the number of observed mutations is known, we can obtain an
estimate of the number of actual mutations by the number that produces
a matching expected number of observed mutations. In the current
setting, the chances of back mutations and multiple mutations on the
same nucleotide are small. Therefore, the difference between the
number of observed mutations and the number of actual mutations is
also very small.

The average synonymous mutation rate $\bar{v}$ (Eq.~\ref{vbar}) comes
into play when we measure the number of mutations $n^{(i)}(t)$ of the
lineage $\alpha^{(i)}$ at time $t$. The expected number of mutations
is then given by 

\begin{equation}
\mathbb{E}[n^{(i)}(t)] = N \bar{v} t,
\label{k1}
\end{equation}
and the expected number of mutations of type $\sigma$ is given by
$N_{\sigma}v_{\sigma} t$. To obtain a probability distribution of
different types of mutations, we assume that mutation events on
\added{different dinucleotide sites are independent,
  allowing for at least a dinucleotide level of site dependence}.

This independent-site assumption is also widely adopted in the
literature \cite{Felsenstein1981nov,Nguyen2014nov,Felsenstein2001oct}
and reflects the stochastic nature of molecular processes and the
feature that DNA replication and proofreading manifest themselves
locally at nucleotide sites.  Under this independence assumption, we
have the following conditional probability:
\begin{align}
  & \mathbb{P}(\boldsymbol{\alpha}^{(i)}= (\alpha_1^{(i)},\ldots,\alpha_n^{(i)})\,\vert\,n^{(i)}(t)=n)
  = \prod_{k=1}^n \frac{N_{\sigma_k}v_{\sigma_k}}{N\bar{v}}, \label{k2}
\end{align}
where $\sigma_k \in \left\{ {\rm +a} , {\rm -a} , {0} \right\}$
defines the type of the $k^{\rm th}$ mutation, which is implicitly
defined by the sequence $(\alpha_1^{(i)},\ldots,\alpha_n^{(i)})$.


Now, let $n_{\rm +a}$ denote the number of synonymous APOBEC3-induced
mutations ($\textrm{TC}\to \textrm{TT}$ and $\textrm{GA}\to
\textrm{AA}$) in the lineage $\alpha$, as exemplified in
Fig.~\ref{fig-schematic}(c). Similarly, we let $n_{\rm -a}$, $n_0$
denote the number of synonymous reverse-APOBEC3 mutations
($\textrm{TT}\to \textrm{TC}$ and $\textrm{AA}\to \textrm{GA}$) and
synonymous non-APOBEC3 mutations, respectively. When the ancestor can
be used as the reference sequence, only the number of APOBEC3-induced
mutations is considered. However, as we discuss in the next section,
when the ancestor sequence is unknown, the forward APOBEC3 mutation on
the test sequence and reverse-APOBEC3 mutations on the reference
sequence can not be distinguished. Conditioned on a total of $n(t)=n$
mutations, the probability of $n_{\rm +a}$ synonymous APOBEC3-induced
mutations arising is
\begin{equation}
  \mathbb{P}(n_{\rm +a}(t) = n_{\rm +a}\mid n(t)=n)
  = \binom{n}{n_{\rm +a}} \left(\frac{N_{\rm +a} v_{\rm +a}}{N\bar{v}}\right)^{n_{\rm +a}}
  \left(1 -\frac{N_{\rm +a} v_{\rm +a}}{N\bar{v}} \right)^{n-n_{\rm +a}}.
    \label{single-binomial}
\end{equation}
In other words, the random variable $n_{\rm +a}(t)$ given $n(t)$
follows a binomial distribution with parameter $n$ and probability
$N_{\rm +a} v_{\rm +a}/(N\bar{v})$.

If the ancestor sequence $S_{0}$ is known, we can infer that for all
samples $\boldsymbol{\alpha}^{(i)}(t^{(i)})$, the number of synonymous
APOBEC3-induced mutations $n^{(i)}_{\rm +a}(t^{(i)})$ follows a
binomial distribution with parameters $n^{(i)}$ and $\frac{N_{\rm +a}
  v_{\rm +a}}{N\bar{v}}$. If $n^{(i)}$ is sufficiently large, we
expect that

\begin{equation}
    {n^{(i)}_{\rm +a}(t^{(i)})}{} \approx \frac{N_{\rm +a} v_{\rm
        +a}}{N\bar{v}} n^{(i)}(t^{(i)}), \quad {\text{for} \quad 1 \ll n^{(i)} \ll N.}
    \label{single-linear}
\end{equation}
Eq.~\eqref{single-linear} reveals a simple linear relationship
followed by all samples.  Unfortunately, we do not know the ancestor
sequence and in order to extract evolution information from data, we
must compare the evolution across multiple lineages.


\paragraph*{Multi-lineage evolution.}
We now consider multiple lineages $\boldsymbol{\alpha}^{(i)}(t^{(i)})$
evolved from the same ancestor sequence $S_0$ and sampled at time
$t^{(i)}$.  The statistics are straightforward if different lineages
evolved independently. However, in reality, some lineages are related
to each other by sharing parts of their evolutionary paths, as
depicted in Fig.~\ref{fig-schematic}(b).

Shared evolutionary paths change how distinct mutations are
enumerated. Suppose that two lineages $\boldsymbol{\alpha}^{(i)}$ and
$\boldsymbol{\alpha}^{(j)}$ share the first $k^{(ij)}$ mutations and
diverge at the $k^{(ij)}+1$-st mutation. After divergence, they
acquire mutations independently.  Among a total of $n^{(i)}(t^{(i)}) +
n^{(j)}(t^{(j)})$ mutations in these two lineages, there are
$k^{(ij)}$ pairs of identical mutations. The subsequent random
mutations are mutually independent of each other.

To construct a systematic notation, for any two lineages
$\boldsymbol{\alpha}^{(i)}$ and $\boldsymbol{\alpha}^{(j)}$, we define
$k^{(ij)}$ to be the largest integer such that $\alpha^{(i)}_\ell =
\alpha^{(j)}_{\ell}$ for all $\ell \leq k^{(ij)}$.  For unrelated
lineages, $k^{(ij)} \equiv 0$. 

Let $A= \big\{ \alpha^{(i)}_{j} \big\}$ denote the set of all
mutations across all lineages. Under our working assumptions, the
$\ell_1^{\rm th}$ mutation $\alpha^{(i)}_{\ell_1}$ in lineage $(i)$
and $\ell_2^{\rm th}$ mutation $\alpha^{(\ell_2)}$ in lineage $(j)$
are identical if and only if they have identical chronological order
($\ell_1 = \ell_2$) in the common evolution history before the two
lineages diverged ($\ell_1 < k^{(ij)}$ and $\ell_2 < k^{(ij)}$). When
the condition $\ell_1=\ell_2 < k^{(ij)}$ is not satisfied, the
mutations are unrelated. Unrelated mutations are independent and
identically distributed. Examples of related mutations $(\alpha_1,
\alpha_2)$ and unrelated mutations $(\alpha_3^{(i)},\alpha_3^{(j)},
\alpha_4^{(i)},\alpha_4^{(j)})$ are shown in
Fig.~\ref{fig-schematic}(b).

Now, pick any two lineages $\alpha^{(i)}$ and $\alpha^{(j)}$, and
define the \textit{relative number of mutations} $n^{(i\mid j)}$ of
$(i)$ with respect to $(j)$ as the number of synonymous mutations
identified when $(j)$ is assumed to be the ancestor sequence,
\textit{i.e.}, when the reference genome is
$(j)$. Fig.~\ref{fig-schematic}(d) provides an example of $n^{(i|j)}$.
Let $t^{(ij)}$ be the time when two lineages
$\boldsymbol{\alpha}^{(i)}$ and $\boldsymbol{\alpha}^{(j)}$ diverged,
as indicated in Fig.~\ref{fig-schematic}(b). Then, $k^{(ij)} =
n^{(i)}(t^{(ij)})=n^{(j)}(t^{(ij)})$.  Under our \textit{infinite
  sites assumption},

\begin{equation}
    \begin{aligned}
        n^{(i\mid j)} & \approx n^{(i)}(t^{(i)})-n^{(i)}(t^{(ij)}) +
        n^{(j)}(t^{(j)}) - n^{(j)}(t^{(ij)}) \\
n_{\rm +a}^{(i\mid j)} & \approx n^{(i)}_{\rm
  +a}(t^{(i)})-n^{(i)}_{{\rm +a}}(t^{(ij)}) + n^{(j)}_{\rm
  -a}(t^{(j)}) - n^{(j)}_{\rm -a}(t^{(ij)}) \\
n_{\rm -a}^{(i\mid j)} & \approx n_{\rm -a}^{(i)}(t^{(i)})-n_{\rm
  -a}^{(i)}(t^{(ij)}) + n_{\rm +a}^{(j)}(t^{(j)}) - n_{\rm
  +a}^{(j)}(t^{(ij)})
    \end{aligned}
    \label{relative-number}
\end{equation}
If $1\ll n^{(i\mid j)}\ll N$, substituting Eq.~\eqref{single-linear}
into the right-hand side of Eq.~\eqref{relative-number}, we find order
of magnitude relationships between the number of mutations between
lineage $(i)$ and reference lineage $(j)$ and the times of sampling
and lineage divergence:

\begin{equation}
    \begin{aligned}
        n^{(i\mid j)} & \approx N \bar{v} \big(t^{(i)} + t^{(j)} -2 t^{(ij)}\big) \\
        n_{\rm +a}^{(i\mid j)} & \approx N_{\rm +a} v_{\rm +a} \big(t^{(i)} - t^{(ij)}\big)
        + N_{\rm -a} v_{\rm -a} \big(t^{(j)} - t^{(ij)}\big) \\
        n_{\rm -a}^{(i\mid j)} & \approx N_{\rm -a} v_{\rm -a} \big(t^{(i)} - t^{(ij)}\big)
        + N_{\rm +a} v_{\rm +a} \big(t^{(j)} - t^{(ij)}\big).
    \end{aligned} 
    \label{multi-asymp}   
\end{equation}
Therefore, upon defining $n_{\rm a}^{(i\mid j)} = n_{\rm +a}^{(i\mid
  j)} - n_{\rm -a}^{(i\mid j)}$, we find the asymptotic expression:

\begin{equation}
    n_{\rm a}^{(i\mid j)} \approx 
    \frac{N_{\rm +a}v_{\rm +a}+ N_{\rm -a}v_{\rm -a}}{N\bar v} n^{(i\mid j)}, 
\quad {\text{for} \quad 1 \ll n^{(i\mid j)} \ll N.}
    \label{multi-linear}
\end{equation}
%

Due to the inability to distinguish APOBEC3-induced mutations on
genome $i$ from reverse-APOBEC3 mutations on genome $j$, the number of
observed APOBEC3-induced mutations $n_{+{\rm a}}^{(i|j)}$ of genome
$i$ with respect to genome $j$ is the sum of $N_{+{\rm a}} v_{+{\rm
    a}}(t^{(i)} -t^{(ij)})$ and $N_{-{\rm a}}v_{-{\rm a}}(t^{(j)} -
t^{(ij)})$. This quantity depends on $(t^{(i)} - t^{(ij)})$ and
$(t^{(j)} - t^{(ij)})$, the branch lengths of both lineage $(i)$ and
lineage $(j)$.  Therefore, the ratio $n_{\rm +a}^{(i|j)}/n^{(i|j)}$
will also depend on the branch lengths and is not a constant when
$N_{+{\rm a}} v_{+ {\rm a}} \neq N_{-{\rm a}}v_{-{\rm a}}$. To avoid
this complication, we must consider both the forward and reverse
mutations, which leads to the desired linearity in
Eq.~\eqref{multi-linear}. Conditioned on $n^{(i|j)} = n$,
$n^{(i|j)}_{\rm a}$ still follows the binomial distribution with mean
and probability parameters $n$ and $p = (N_{\rm +a}v_{\rm +a}+ N_{\rm
  -a}v_{\rm -a})/(N\bar{v})$:
\begin{equation}
  \begin{aligned}
    &\mathbb{P}\big(n_{\rm a}^{(i|j)}(t^{(i)}) = n_{\rm a}^{(i\mid
  j)} \mid n^{(i\mid j)} = n\big) \\ & \hspace{2cm}=  \binom{n}{n_{\rm a}^{(i\mid
    j)}}
    \left(\frac{N_{\rm +a}v_{\rm +a}+ N_{\rm -a}v_{\rm -a}}{N\bar{v}}\right)^{n_{\rm a}^{(i\mid j)}}
    \left(1-\frac{N_{\rm +a}v_{\rm +a} + N_{\rm -a}v_{\rm -a}}{N\bar{v}}\right)^{n-n_{\rm a}^{(i\mid j)}}.
  \end{aligned}
    \label{multi-binomial}
\end{equation}

When an \textit{unrelated} genome $\ell$ is used as a reference, the
proportionality indicated in Eq.~\eqref{multi-linear} is no longer
valid. In this case, because of unrelatedness and the infinite sites
assumption, we have $n^{(i|\ell)} = n^{(i|j)} + \Delta n^{(j|\ell)}$
and $n_{\textrm{a}}^{(i|\ell)} = n_{\textrm{a},i|j} + \Delta
n_{\textrm{a}}^{(j|\ell)}$, where $\Delta n^{(j|\ell)}$ and $\Delta
n_{\textrm{a}}^{(j|\ell)}$ are two values that are independent of
$i$. Consequently, $n_{\textrm{a}}^{(i|\ell)}$ and $n^{(i|\ell)}$ are
not proportional to each other but follow the linear relationship

\begin{equation}
  n_{\textrm{a}}^{(i|\ell)} \approx \frac{N_{\rm +a}v_{\rm +a}+ N_{\rm -a}v_{\rm -a}}{N\bar v} n^{(i|\ell)} 
+ \left[\Delta n_{\textrm{a}}^{(j|\ell)} - \frac{N_{\rm +a}v_{\rm +a}+ N_{\rm -a}v_{\rm -a}}{N\bar v}
\Delta n^{(j|\ell)}\right].
  \label{multi-linear2}
\end{equation}
The fraction $({N_{\rm +a}v_{\rm +a}+ N_{\rm -a}v_{\rm -a}})/({N\bar
  v})$ and constant $\Delta n_{\textrm{a}}^{(j|\ell)} -({N_{\rm
    +a}v_{\rm +a}+ N_{\rm -a}v_{\rm -a}})/({N\bar v})$ are parameters
to be obtained from data. If genomes $(j)$ and $(\ell)$ have evolved
from the same ancestor with the same mutation rates, the constant term
is zero. Nonzero constant term partially measures the degree to which
the mutation rates of the two lineages are different.

\paragraph*{Relaxing \replaced{relative clock}{molecular clock}
  assumptions.} \replaced{So far in our relative clock
  construction}{sections}, we assumed that the raw mutation rates are
constant over genomic locations, lineages, and
time. \added{Although this assumption is not realistic,
  phylogenetic-like models do not require specification of raw
  mutation rates. Since our hybrid approach shares features with
  phylogenetic methods, we can relax the constant mutation rate
  assumption while retaining key relative rate information. The
  assumption of constant mutation rates across lineages and
  generations can be relaxed by properly reinterpreting the time
  variable $t$, while the assumption of constant mutation rates across
  genomic locations can be relaxed by changing how total mutation
  rates are calculated.}

For time-varying overall mutation rates, $v(t)$, we define a new
nonlinear measure of time $\tau \coloneqq \int_0^t v(t)\mathrm{d} t$,
over which the mutation rate is again constant.  Interpreted
biologically, $\tau$ is interpreted biologically as being proportional
to the number of DNA replication cycles since the last common
ancestor. It has been found that the mutation rate is positively
correlated with DNA replication frequency, and negatively correlated
with the generation time
\cite{Bromham2003mar,Li1996feb,Moorjani2016sep}.  Under those
circumstances, we replace the absolute time $t$ in the previous
sections with a pseudo-time $\tau$ that measures the number of
generations since the last common ancestor, which is universal across
all different lineages. Since Eqs.~\eqref{single-linear} and
\eqref{multi-linear} are independent of the time variable, they are
invariant for the new time $\tau$.

Incorporating possible heterogeneity in mutation rates across
different sites requires a more complex argument.  Genomic sites with
an overall higher mutation rate contribute most significantly to the
mutations observed between lineages.  Instead of simply multiplying
$v_{\sigma_k}$ by the number of sites $N_{\sigma_k}$ available for the
mutation of type $\sigma_{k}$, we need to calculate the total mutation
rate of type $\sigma_k$ via a sum
$\sum_{k=1}^{N_{\sigma_k}}v_{\sigma_k,s}$ that weights the mutation
rates at each site $s$. As the mutations accumulate, indirect and
complex cellular interactions may also change the rate of mutation.
However, under an infinite sites approximation, only a very small
fraction of the sites have mutated. Therefore, the total mutation
rates and their ratios remain constant for a very long time. Only in
extreme cases where very few sites have extremely high mutation rates
while all other sites have negligible mutation rates, is the infinite
sites assumption effectively violated and ratios of the \textit{total}
mutation rates can no longer be assumed constant.  In light of this
argument, we conclude that the spatial homogeneity assumption can be
effectively relaxed provided the distribution of mutation rates are
not too disperse. After relaxation of these assumptions and proper
reinterpretation, our previous calculations and equations hold.

\paragraph*{Statistical tests for changes in the mutation rate.}
Although time-dependent changes in \textit{separable} mutation rates
$v(t)$ can be taken into account and do not affect mutation rate
ratios, our modeling approach allows one to statistically test for
inseparable shifts in the mutation rate where ratios of mutation rates
change. A model in which mutation rates change in time is simulated in
Appendix ``Hypothetical scenarios'' and shown in
Fig.~\ref{fig:resp_2}(a).  Specifically, Eq.~\eqref{multi-binomial}
(the distribution of the count of a specific mutation type,
conditioned on known relative mutation rates and the total number of
synonymous mutations) enables the calculation of a 95\% prediction
interval. Instances where samples fall outside this interval suggest a
significant alteration in the mutation rate. We provide an example of
this test in the Results and in Fig.~\ref{fig:resp_2}(b).
%

Eq.~\eqref{multi-binomial} can be further used to test whether a focal
genome and a set of known genomes share the same relative mutation
rates. To enhance resolution, the focal genome is used as a reference
and its coordinates are to the origin $(n, n_{\rm a})=(0,0)$. The null
hypothesis that reference genome and the set of other genomes $(i)$
exhibit the same relative mutation rate can be tested by how well
$(n^{(i)}, n_{\rm a}^{(i)})$ represent draws from the binomial
distribution given in Eq.~\eqref{multi-binomial}. For example, one can
can apply least-squares regression on the data points $(n^{(i)},
n_{\rm a}^{(i)})$ for each genome $i$. Under the null hypothesis, the
$n_{\rm a}$-intercept of the regression line should be at $b \equiv
n_{\rm a}(n=0)=0$. A $t$-test statistic given by $b/\textrm{SE}(b)$
can then be used, providing a sensitive test of the null hypothesis.


\subsection*{Selection effects and model validation via simulation}
Our theoretical analysis predicts a linear relationship between
$n_{\textrm{a}}^{(i|j)}$ and $n^{(i|j)}$, which we test
by implementing two types of simulations.

\paragraph*{Mutation-only processes.}
First, we simulate DNA evolution using independent and identically
distributed (i.i.d.) GTR processes.  For simplicity, we first exclude
selection pressure and assume equal frequencies and equal mutation
rates of all four nucleotides, aligning our model with the JC69 limit
\cite{JUKES1969}.
\begin{figure}[htbp!]
  \centering 
  \includegraphics[width = 5in]{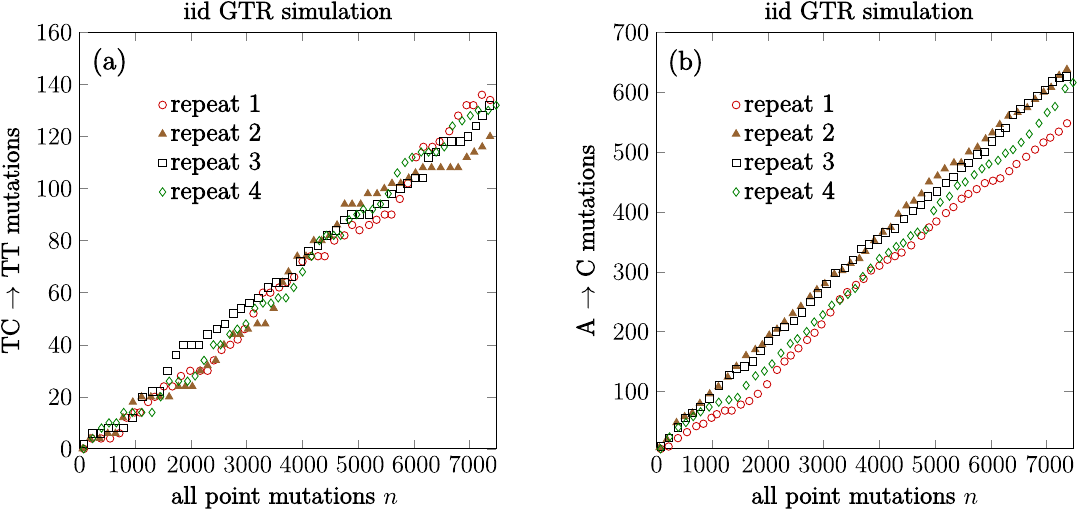}
  \caption{\added{Simulations of different scenarios. (a-b) Four
      representative trajectories of an iid substitution process of
      the JC69 limit.  A total of $10^6$ nucleotides are used for each
      simulation and both TC $\to$ TT and $\textrm{A}\to \textrm{C}$
      mutations are linearly correlated with the total number of point
      mutations.}}
  \label{fig:sim_1ab}
\end{figure}

For each trial, we randomly sample $10^6$ nucleotides from the
equilibrium distribution of $\{$A,T,C,G$\}$ to construct an initial
sequence. For each nucleotide position, we then simulate a Markov
mutation process using the JC69 transition matrix and a variant of the
Gillespie algorithm~\cite{Gillespie1977dec}. During the simulation of
the substitution model, we count the total number of mutations and
plot the associated numbers of TC $\to$ TT and A $\to$ C mutations in
Fig.~\ref{fig:sim_1ab}(a) and (b), respectively.  Across all four
independent trajectories shown, the linear relationship between the
number of all mutation types and the total mutation number holds, even
up to $\sim 10^{4}$ mutations.  This justifies and implies broad
applicability of the infinite sites assumption.

\paragraph*{Mutation-selection processes.}
In a second set of simulations, we factor in selection pressure, using
a simple asexual reproduction logistic growth model with a shared
carrying capacity $K$.  In this simulation, track the population of
genes through the numbers of background synonymous mutations $n -
n_{\textrm{a}}$, synonymous APOBEC3-driven mutations $n_{\textrm{a}}$,
and ``hidden'' (nonsynonymous) beneficial mutations $n_{\textrm{h}}$.
The population undergoes a birth-death process
defined by birth and death rates 
\begin{align}
  \beta(n_{\rm a}, n_{\rm h}) &  = \beta_0 s(n_{\rm a}, n_{\rm h}), \label{eq:birth}\\
  \mu(n_{\rm a}, n_{\rm h}) &= \mu_0 + \beta(n_{\rm a}, n_{\rm h}) \frac{N_{\rm G}(t)}{K},
  \label{eq:death}
\end{align}
where $N_{\rm G}(t)$ is the total population of genomes, and $K$ is the
carrying capacity, and the selection coefficient is given by 
\begin{equation}
  s(n_{\rm a}, n_{\rm h}) = 1 + \sigma_{\rm a} n_{\rm a} + \sigma_{\rm h} n_{\rm h}.
  \label{eq:fitness}
\end{equation}
The selection coefficient depends linearly on the numbers of mutations
$n_{\rm a}$ and $n_{\rm h}$ through the selection strengths and
$\sigma_{\rm a}$ and $\sigma_{\rm h}$.  Genomes that have larger
$s(n_{\rm a}, n_{\rm h})$ reproduce faster.
\begin{figure}[h!]
  \centering 
  \includegraphics[width = 5.4in]{fig_resp_1c-f.pdf}
  \caption{\added{Simulations of scenarios that include
      mutation and selection following
      Eqs.~\ref{eq:birth}-\ref{eq:fitness}. We used $\beta_{0}=1$,
      $\mu_{0}=0.5$ and assumed that with each birth, one daughter has
      equal probability 0.1 of acquiring each type of mutation.  The
      total initial numbers of possible mutations $N_{i}$ for
      synonymous non-APOBEC3 mutations, synonymous APOBEC3 mutations,
      and hidden nonsynonymous mutations is set to $10^{5}, 10^{3}$,
      and $10^{3}$, respectively.  (a) Numbers of synonymous
      APOBEC3-relevant mutations $n_{\rm a}$ associated with the
      numbers of synonymous non-APOBEC3-relevant mutations $n - n_{\rm
        a}$ from samples of simulation trajectories.  Here, no
      selection for hidden mutations ($\sigma_{\rm h}=0$) is
      present. (b) Each diamond symbol represents the slope of an
      independently simulated selection-mutation process, as shown on
      (a). Slopes for different $\sigma_{\rm a}$ and 25 uniformly
      sampled values of $\log K$ are shown.  (a) and (b) show that
      large population size $K$ and strong selection coefficient
      $\sigma_{\rm a}$ can lead to a larger effective mutation rate of
      APOBEC3-driven mutations. By design, under weak selection
      strength $\sigma_{\rm a}$, the effective mutation rate of
      APOBEC3-driven mutations is identical to the non-APOBEC3-driven
      mutations. (c) Values of $n_{\rm a}$ associated with $n - n_{\rm
        a}$ under different strengths $\sigma_{\rm h}$ of selection of
      hidden beneficial mutations. Strong selection of
      hidden beneficial mutations can restore the increased effective
      mutation rate of APOBEC3-driven mutations to the same level as
      the pure mutation processes. (d) For large $\sigma_{\rm h}>
      \sigma_{\rm a}$, the expansion of $n_{\rm a}$ is limited (red
      diamonds).}}
  \label{fig:sim_1cf}
\end{figure}

At birth, one daughter has a probability to acquire an additional
mutation of specified type.  Mutation probabilities are chosen such
that the total rates of each type of mutation are identical.  To
connect with the real-world evolutionary process, after each unit of
time (corresponding to the average time between replication of a
single genome), a sample of the population is drawn and genomes with
specific number $n-n_{\rm a}$ of synonymous non-APOBEC3 mutations
$n-n_{\rm a}$ are collected, along with their associated numbers of
synonymous APOBEC3 mutations $n_{\rm a}$.  These $(n_{\rm a}, n-n_{\rm
  a})$ points are collected across all sampled time points and plotted
in Figs.~\ref{fig:sim_1cf}(a) and (c). From the infinite sites
approximation, Eq.~\ref{k1}, and the constant relationship between
different types of mutation, $n-n_{\rm a}$ is tightly associated with
time.

%

%
%


\added{We assign different selection coefficients $\sigma_{\rm a}$ and
$\sigma_{\rm h}$ to understand how selection on the synonymous APOBEC3
mutations and other beneficial mutations may alter the linear
relationship between $n_{\rm a}$ and $n$. In a mutation-only process
(negligible $\sigma_{\rm a}$ and $\sigma_{\rm h}$), the expected
number of mutations of each type is the same as the others,
\textit{e.g.}, $n_{\rm a} \approx (n - n_{\rm a})$, as indicated by
the blue trajectory in Fig.~\ref{fig:sim_1cf}(a).  In
Fig.~\ref{fig:sim_1cf}(b), we plot, for each value of $\sigma_{\rm a}$
indicated, the trajectory slopes for 25 different values of $\log_{10} K$
uniformly sampled within $\log_{10} K \in [2,4]$ or $K \in [10^2, 10^4]$. The increased
mutation rate with larger population size is consistent with prior
theoretical understanding of selection effects
\cite{lewontin2000evolutionary,Ingvarsson2008}. Even when the
selection for APOBEC3 mutations is strong ($\sigma_{\textrm{a}} \geq
0.01$) and the carrying capacity is large ($K > 10^3$), the effective
APOBEC3 mutation rate is at most doubled, as shown in
Fig.~\ref{fig:sim_1cf}(b).}

\added{If the selection of the hidden beneficial mutations is large
  (say, $\sigma_{\rm h} \gtrsim \sigma_{\rm a}$), the ratio $n_{\rm
    a}/(n - n_{\rm a}) \approx 1$, as shown by the red trajectory in
  Fig.~\ref{fig:sim_1cf}(c) and the marked red values of slopes (diamonds) in
  Fig.~\ref{fig:sim_1cf}(d). Here, the hidden beneficial mutations
  drive selection, leading again to comparable $n_{\rm a}$ and
  $n-n_{\rm a}$.}

\added{Summarizing, our simulations verify (i) that the infinite sites
  hypothesis holds up to $\sim 10^3$ mutations in a genome of $10^6$
  bp, (ii) that the linear relationship between the number of
  different types of mutations holds even under selection, (iii) that
  the effective mutation rate of APOBEC3-driven mutations can be
  increased by selection only if they are relatively strongly selected
  for ($\sigma_{\rm a} \geq 0.01$), and (iv) that strong selection for
  other beneficial mutations can mask the increase in the effective
  mutation rate of APOBEC3-driven mutations due to selection.}

\subsection*{Sequence analysis and alignments}
We obtained the sequences of 237 MPXV genomes from the NCBI database
as listed in Table \ref{table1}. To date, there are hundreds of MPXV
genomes available. Dates of the sequences are inferred from the
collection date and other contextual information. However, a majority
of them are samples collected from the most recent outbreak. Thus, a
large number of them are closely related, and using our proposed
analysis, contribute little information to the evolutionary history of
MPXV.  The sequences were aligned using the affine gap model from the
\texttt{BioAlignments} package \cite{bioalignments} in
\texttt{Julialang} \cite{Julia-2017}. Alignment results are
manipulated using the \texttt{BioSequences} and
\texttt{GenomicAnnotations} packages in \texttt{Julialang}.

\added{The dinucleotide and synonymous mutations were identified by
  first enumerating all mutations in the alignment, in the context of
  whole genome sequences. Then, we identify the synonymous mutations
  by comparing the amino acid sequences of the genes. These
  considerations are not restricted to the third codon position.}

For sequences without gene annotations, we used NC\_063383 as the 
template to annotate the genes. Then, these annotations were used to 
generate pairwise alignment results and identify synonymous mutations.
The statistical tests were developed based on likelihood ratio tests
for binomial distributions. 

Appendix {\bf Lineage Analysis} provides details on the inference of a
tree based on ordering of genome mutations. Our result is shown in
Fig.~\ref{fig:result_3}(a) and compares well with the phylogenetic
tree shown in Fig.~\ref{fig:result_3}(b) which as derived from using
standard NCBI BLAST \cite{King2010}. However, BLAST does not correctly
resolve subtle ancestral relationships between recently collected
genomes, which our method does.

\section*{Results}
\subsection*{Statistically distinct subgroups with different
evolutionary features}\label{sec:evol-landscapes}
For the reasons discussed in {\bf Multi-lineage evolution},
\added{when the phylogenetic tree is not resolved, we
  cannot distinguish APOBEC3-induced mutations from their reverse
  mutation. If  the phylogenetic tree is established, these two
  directions of substitutions can be distinguished from established
  methods such as the UNREST model\cite{Yang1994jul}.} Here,
APOBEC3-relevant mutations are defined as the mutations matching
patterns $\textrm{TC}\leftrightarrow \textrm{TT}$ or
$\textrm{GA}\leftrightarrow \textrm{AA}$, as detailed in Table
\ref{table:names}. The number of these synonymous mutations is
$n_{\textrm{a}}=n_{+\textrm{a}}+n_{-\textrm{a}}$.
\added{For a time-scaled phylogenetic tree construction
  method, ``root-to-tip'' regression is used to estimate the
  evolutionary rate of the virus, represented by the slope of the
  regression line\cite{Rambaut2016jan}. In the absence of a resolved
  phylogenetic tree, a linear fit of two types of mutations in our
  relative clock is essentially a ``tip-to-tip'' regression.}

\begin{figure}[h!]
  \centering
  \includegraphics[width=5.2in]{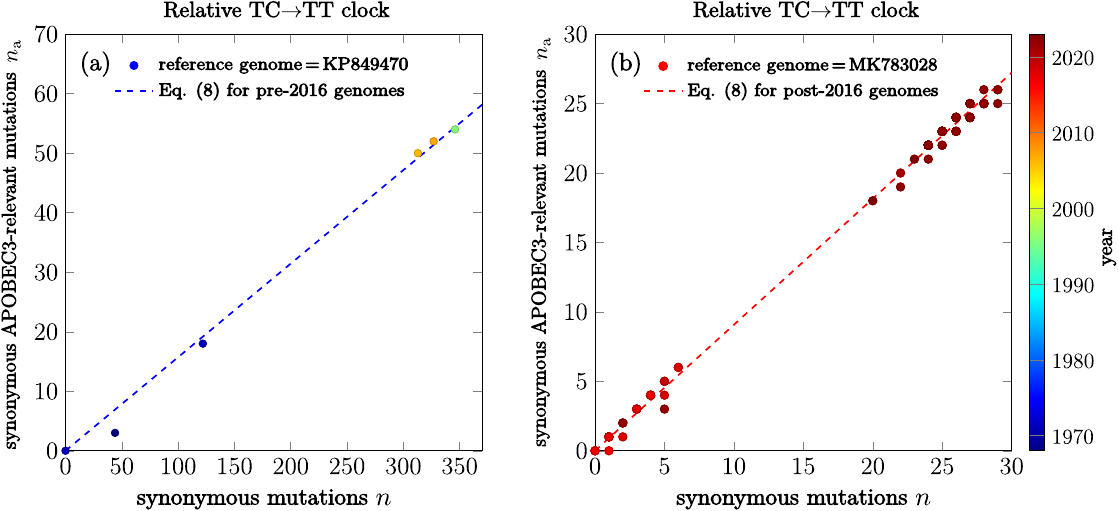}
  \caption{Relative molecular clocks for different types of
    mutations with respect to an arbitrarily chosen reference genome,
    instead of time, exhibit linearity. (a) The number of synonymous
    APOBEC3-relevant mutations (TC $\to$ TT) is plotted against the
    number of synonymous mutations for the genomes collected before
    2016. (b) The number of synonymous APOBEC3-relevant mutations (TC
    $\to$ TT) is plotted against the number of synonymous mutations
    for the genomes collected after 2016.}
  \label{fig:result_1}
\end{figure}

\paragraph{Successful application of the \replaced{relative}{recalibrated} 
molecular clock to pre-2016 genomes.} We obtained 6 MPXV genomes
collected before 2016 from GenBank. The genome KP849470 was
arbitrarily chosen as the reference genome against which other genomes
are aligned.  For each genome $(i)$, we calculate the relative number
of synonymous mutations $n^{(i\mid \textrm{KP849470})}$ and number of
APOBEC3-relevant mutations $n_{\rm a}^{(i\mid \textrm{KP849470})}$.
For simplicity, we omit the reference genome in the superscript and
write $n^{(i)}$ and $n^{(i)}_a$ respectively.
Each genome is then plotted on the $(n, n_{{\rm a}})$ plane, where they
fall near a line which we find by using a least-squares fit and
Eq.~\eqref{multi-linear}. As shown in Fig.~\ref{fig:result_1}(a), the
data points are well fitted by the linear model, allowing us to
conclude the pre-2016 MPXV genomes have evolved with a constant set of
relative mutation rates.

Given the shared relative mutation rates, we can use the linear fit as
a baseline to compare the APOBEC3-relevant mutation rates. A total of
$N \sim 100,000$ possible synonymous mutations were identified on the
reference genome KP849470, with $N_{+\rm a}\sim 3,000$ of them
APOBEC3-induced, and $N_{-\rm a}\sim 8,000$ of them
reverse-APOBEC3-induced. In total, APOBEC3-relevant mutations account
for about 11\% of all possible synonymous mutations. By contrast, the
slope of the linear fit in Fig.~\ref{fig:result_1}(a) is $0.157 \pm
0.007$, which represents the fraction of observed APOBEC3-relevant
mutations ${n_{\rm a}}/{n}$. This fraction is significantly larger
than the fraction of possible APOBEC3-relevant mutations $N_{\rm
  a}/N\approx 0.11$, suggesting that the APOBEC3-relevant mutation
rate is higher than that of the average mutation already extant in the
pre-2016 MPXV genomes. The ratio $n_{\rm a}^{(i)}/n \sim 0.157$ serves
as a better baseline to test whether the APOBEC3-relevant mutation
rate is higher in the post-2016 MPXV genomes.

\paragraph{Successful application of the 
\replaced{relative}{recalibrated} molecular clock
to post-2016 genomes.} We obtained 226 MPXV genomes from GenBank
collected after 2016. Genome MK783028 was chosen to be the reference
genome against which other genomes are aligned.  The relative number
of synonymous mutations $n^{(i\mid \textrm{MK783028})}$ and number of
APOBEC3-relevant mutations $n_{\rm a}^{(i\mid \textrm{MK783028})}$ are
computed and each genome is plotted in the $(n, n_{{\rm a}})$
plane. The results are shown in Fig.~\ref{fig:result_1}(b). These
points also fall near a line, indicating that the post-2016 MPXV
genomes also evolved with a constant set of relative mutation rates.

The total number of possible mutations of different types on the
reference genome MK783028 is similar to that of KP849470. However, the
slope $n_{\rm a}/n$ of the linear fit in Fig.~\ref{fig:result_1}(b) is
$0.912 \pm 0.002$ which is significantly larger than the baseline
slope $0.157 \pm 0.007$ of the pre-2016 genomes.

\paragraph{Interpretation of different APOBEC3-relevant mutation rates.}
In conclusion, our key assumption of constant ratios of mutation rates is
separately applicable to the two subgroups of genomes (distinguished
by their year of collection). The mutation rate ratios are particular
to each group. Because we have factored out the selection pressure by
considering only synonymous mutations, selection is unlikely to
account for the difference in the two mutation rate ratios. If this
difference is due to random neutral mutations within the animal hosts,
we would expect that the ratio of the number of APOBEC3-relevant
mutations to the number of all synonymous mutations is the same for
both groups of genomes resulting in similar slope in
Fig.~\ref{fig:result_1}.  In addition, since the total number of
possible mutations of different types on the reference genome is
similar for both groups of genomes, the difference in the observed
frequencies of APOBEC3-relevant mutations can only be explained by the
difference in the relative APOBEC3-relevant mutation rate.

The SNP variants across the pre-2016 genomes \added{that} often belong
to different clades likely reflect the evolution of the virus in
unknown animal reservoirs. By contrast, the genomes in the post-2016
group are clustered into a single clade, and the associated SNP
variants likely reflect the evolution of the virus in the human host.
Therefore, we conclude that the APOBEC3-relevant mutation frequency
$n_{\rm a}/N_{\rm a} > n/N$ is \textit{relatively} higher in the
pre-2016 group ($n_{\rm a}/N_{\rm a} > n/N$), possibly due to some
intrinsic or extrinsic physical or chemical mutagens like UV
radiation, as mentioned previously.  The APOBEC3-relevant mutation
rate is extremely high in the post-2016 group, which could be a result
of persistent APOBEC3-editing in the human host.

\added{APOBEC3 and other human-specific environmental factors may also
  impose selection pressure that favors certain mutations. For
  example, tRNA abundance may be different in the human host compared
  to the animal reservoirs. However, such effects are generally
  expected to be gene-specific, exhibiting a stronger effect on highly
  expressed genes. Our analysis in Appendix ``Distribution of
  APOBEC3-induced mutations'' does not support this hypothesis as we
  do no find exceptionally high APOBEC3-relevant per-site mutation
  rates in specific genes (see Fig.~\ref{fig:supp_3}).  Neither are
  mutation types correlated across genes (see
  Fig.~\ref{fig:supp_4}). Moreover, our simulations show that even a
  strong selection coefficient favoring APOBEC3-relevant mutations can
  at most double the APOBEC3-relevant mutation rate, which is much
  smaller than the observed difference between the pre-2016 and
  post-2016 groups (see Fig.~\ref{fig:sim_1cf}). Other beneficial
  nonsynonymous mutations could further reduce the increase in the
  ratio.  Therefore, we conclude that the observed difference in
  APOBEC3-relevant mutation rate is mostly likely due to a difference
  between the APOBEC3-editing activity in animal reservoirs and the
  human host.}

\paragraph*{General applicability of \replaced{relative}{recalibrated} 
molecular clocks and bias for APOBEC3-relevant mutations.}  We have
shown that our recalibrated molecular clock method is applicable to
both pre- and post-2016 genomes in the context of APOBEC3-relevant
mutations versus all synonymous mutations. In fact, the
\replaced{relative}{recalibrated} molecular clock method
is applicable to other types of synonymous mutations as well.  For
example, we considered the A $\to$ C type of synonymous mutations
versus synonymous non-APOBEC3-relevant mutations in
Fig.~\ref{fig:supp_1}. The linearity is still observed
and is consistent in both pre- and post-2016 genomes.  This
observation suggests not only that the
\replaced{relative}{recalibrated} molecular clock method
is generally applicable\deleted{, but that APOBEC3-relevant
  mutations in the post-2016 genomes are much more frequent than other
  types of synonymous mutations, including ``A $\to$ C'' mutations}.

\subsection*{The ancestor of post-2016 MPXV genomes}\label{sec:origin}
\paragraph{The geometry of pre- and post-2016 genomes on the same plane.}
By choosing a common reference genome for the two groups of genomes,
we can visualize them on the same $(n, n_{\rm a})$-plane. The 
geometric relationship between the two groups of genomes reveal information 
about the evolutionary history of the virus in these two groups of hosts.

\begin{figure}[htbp]
  \centering
  \includegraphics[width=\textwidth]{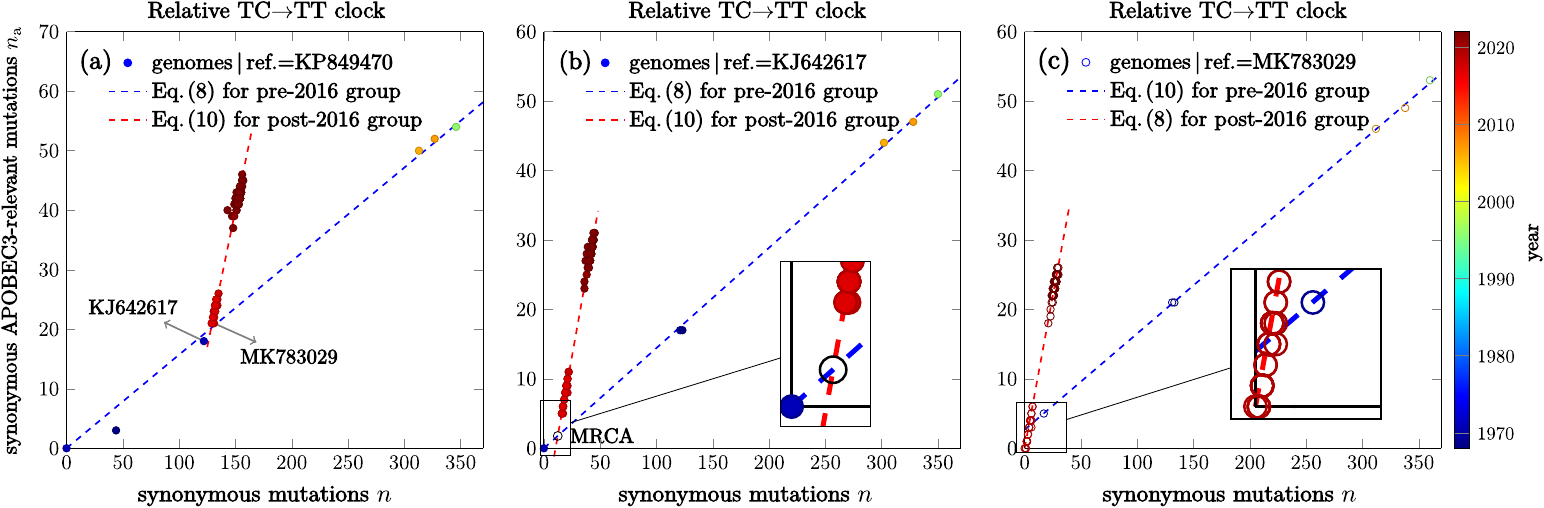}
  \caption{(a) The number of observed APOBEC3-relevant mutations
    $n_{\rm a}$ of all genomes plotted against the number of observed
    synonymous mutations $n$, using the genome KP849470 as a
    reference. The two genomes KJ642617 and MK783029 are highlighted
    since they are the closest genomes to the intersection of the two
    lines corresponding to the pre-2016 and post-2016 genomes. (b) The
    number of observed APOBEC3-relevant mutations $n_{\rm a}$, using
    genome KJ642617 as the reference, plotted against the number of
    observed synonymous mutations $n$. The open circle represents the
    location of the hypothesized most recent common ancestor
    (\textit{MRCA}) of the post-2016 genomes in the animal reservoir.
    The inset shows that the line of the post-2016 group does not
    intersect the origin, suggesting that KJ642617 is not a common
    ancestor of the post-2016 group. (c) The number of observed
    APOBEC3-relevant mutations $n_{\rm a}$ plotted against the number
    of observed synonymous mutations $n$, using MK783029 as reference.
    The inset shows that the line of the pre-2016 group does not pass
    through the origin, suggesting that MK783029 has already undergone
    some APOBEC3-editing after its ancestor's transmission to human.}
  \label{fig:result_2}
\end{figure}

When KP849470 is used as the reference genome, the post-2016 genomes
are still distributed along a line in the $(n, n_{\rm a})$-plane, as
shown in Fig.~\ref{fig:result_2}(a). However, since the post-2016
genomes have undergone a different evolutionary history, the genome
KP849470 should not be co-linear with the post-2016 genomes. In other
words, the line representing the post-2016 genomes should not pass
through the origin, but the lines of the pre- and post-2016 genomes
will intersect at some point. Genomes at this intersection will share
a common evolutionary history in both the animal reservoir and the
human host. \added{Since the post-2016 genomes may come from the
  animal reservoir, the intersection point represents the most recent
  common ancestor (MRCA) of the post-2016 genomes in the animal
  reservoir.}

The genomes KJ642617 and MK783029 lie approximately at the
intersection of the two lines when the genome KP849470 is used as
reference. They are thus \deleted{are} good candidates for the MRCA of
the post-2016 genomes. Small deviations from the precise intersection
point of the two lines can be explained by the stochastic nature of
mutations.

\paragraph{Statistical tests to determine the MRCA of the post-2016 MPXV genomes.}
\added{In order to minimize the effects of stochasticity and improve
precision, we determine whether genomes KJ642617 and MK783029 are the
MRCA of interest by choosing them as the reference genome (setting
their positions to $(n, n_{\rm a})=(0,0)$), respectively. Next, we
determine how well the two lines corresponding to the pre-2016 and
post-2016 genomes intersect at the origin.}

\added{Using least-squares regression, we found that when KJ642617 is
  used as the reference genome (Fig.~\ref{fig:result_2}(b)), the
  $n$-intercept of the linear fit of the post-2016 genomes is $-9.27
  \pm 0.34$ but when MK783029 is used as reference
  (Fig.~\ref{fig:result_2}(c)), the $n$-intercept of the pre-2016
  genomes is $2.80 \pm 0.55$.  In both cases, the p-values of the null
  hypothesis that the $n$-intercept is zero are less than 0.01.  Thus,
  neither KJ642617 nor MK783029 is representative of the MRCA of the
  post-2016 MPXV genomes.}

\added{However, when we apply the combined p-value test described
  previously, based on the conditional probability formula
  Eq.~\eqref{multi-binomial}, the p-values for the null hypotheses are
  both greater than 0.05. This discrepancy reflects the fact that the
  observed number of APOBEC3-relevant mutations has a sharper
  distribution concentrated around the mean value than the
  distribution predicted in theory. This sharper distribution might
  result from multiple factors, including the close phylogenetic
  relationship between genomes and selection pressure.}

\deleted{Given that we have
a binomial probability model for the distribution of the number of
APOBEC3-relevant mutations given the total number of synonymous
mutations, we can also use a likelihood ratio test based on
Eq.~\eqref{multi-binomial}. This test is conceptually better than an
ordinary least-squares regression, as the latter requires the error
terms to be independent and normally distributed with identical
variance.}

\deleted{To account for dependence between closely related genomes, we sample
one genome from all genomes collected in the same year, and assume
that the genomes collected in different years are independent. We then
resample 1000 times to obtain an average of the p-values of the null
hypothesis that the sampled genomes follow Eq.~\eqref{multi-binomial}
with respect to the reference genome of interest. The alternative
hypothesis is that they follow the distribution with a nonzero offset
similar to that in Eq.~\eqref{multi-linear2}. This null hypothesis is
a rigorous formulation of the intuitive idea that these genomes fall
on the same line in the $(n, n_{\rm a})$-plane. When this null
hypothesis is rejected, we conclude that the reference genome and the
sampled genomes have undergone evolution with different relative
mutation rates.}

\deleted{For genome KJ642617, we reject, with p-value 0.00092, the null
hypothesis that post-2016 genomes follow the distribution specified by
Eq.~\eqref{multi-binomial} and conclude that KJ642617 is not the MRCA
of the post-2016 MPXV genomes.}

\deleted{For genome MK783029, the average p-value of the null hypothesis that
pre-2016 genomes follow the distribution specified by
Eq.~\eqref{multi-binomial} is 0.070, which is not statistically
significant. This is mostly because we have a very limited number of
pre-2016 genomes. In addition, all the observed APOBEC3-relevant
mutations of the MK783029 genome with respect to KJ642617 are of the
APOBEC3-induced type. This high fraction of APOBEC3-induced mutations
in APOBEC3-relevant mutations suggests that the MK783029 genome has
undergone several APOBEC3 editing events since the MRCA's entry into
human hosts. By contrast, the number of APOBEC3-induced mutations and
the number of reverse-APOBEC3 mutations of other pre-2016 genomes are
comparable, as shown in Fig.~\ref{fig:supp_2}.  Therefore, we
conclude that MK783029 is not representative of the MRCA in the animal
host.}



%

\paragraph{Statistical inference of the MRCA in the animal host.}
Although neither KJ642617 nor MK783029 are good candidates for the
MRCA in the animal hosts, we can infer the number of synonymous
mutations with respect to a given reference genome by analyzing the
intersection of the linear fits to the pre- and post-2016 genomes.

To estimate the time at which the MRCA first emerges, the reference
genome should be chosen from the pre-2016 group. Only if the reference
genome is chosen from the pre-2016 group will the critical ancestor
have a fewer number of synonymous mutations than other post-2016
genomes. Such a choice would allow us to linearly extrapolate the strict
molecular clock fit to the post-2016 genomes and find the emergence
time of the MRCA.

\added{Although we can use any pre-2016 genome as a reference,
  KJ642617 is the most similar to the post-2016 genomes, as measured
  by the number of observed synonymous mutations.  For mutations that
  occur in a Poisson-like process, the effects of stochasticity
  measured by standard deviation is proportional to the square root of
  total number of mutations.  Therefore, using KJ642617 as the
  reference reduces the number of observed mutations and therefore the
  noise, providing a more precise estimate. As shown in
  Fig.~\ref{fig:result_2}(a), ordinary least-squares regression
  provides an estimate of the 95\% confidence interval
  $n_{\textrm{ancestor}} \in [11.61,12.76]$ of the number of
  synonymous mutations of the critical ancestor with respect to
  KJ642617.}

\subsection*{Molecular clock of the post-2016 group}

The knowledge of the relative number of synonymous mutations of the
critical ancestor is helpful for timing its emergence assuming a strict
molecular clock is present during its subsequent evolution.
%
  \begin{figure}[htbp]
  \begin{center}
    \includegraphics[width=2.8in]{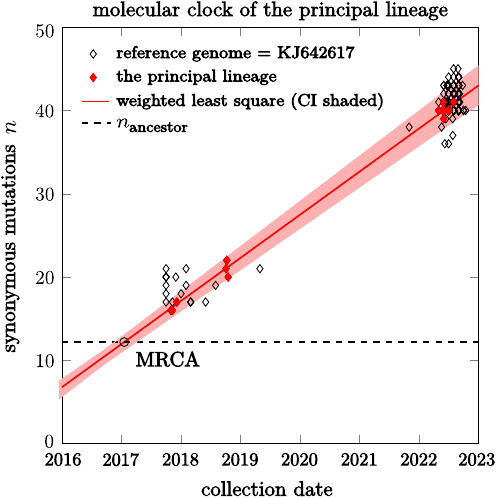}
    \end{center}
  \caption{The number of synonymous APOBEC3-induced mutations plotted
    against the collection date of the post-2016 genomes, using
    KJ642617 as the reference. The red diamond represents the genomes
    within the lineage with the longest branch length, starting with
    MK783028 and ending with ON694329.}
  \label{fig:result_clock}
\end{figure}


In order to understand whether mutation rates have changed over time,
\replaced{we apply in Appendix {\bf Lineage analysis} the root-to-tip
  regression, using the genetic distance defined by the number of
  synonymous mutations and calibration data defined by manually
  identified collection dates of the genomes, over a phylogenetic tree
  constructed by asymmetry of TC $\to$ TT mutations. The total rate
  $N\bar{v}$ of synonymous mutations for the post-2016 genomes is
  estimated to be around 5 per year.}{we need the collection times of
  each genome, which we manually identified.} Their number of
synonymous mutations $n^{(i)}(t^{(i)})$ are plotted against the
collection \replaced{date}{year}, as shown in
Fig.~\ref{fig:result_clock}.

Multiple factors such as differences between emergence and collection
times, cross-lineage and cross-generation differences in the mutation
rates, and different geographic locations can result in deviations
from the linear relationship between the number of mutations and the
collection time predicted by the strict molecular clock.
Consequently, unlike the
\replaced{relative}{recalibrated} clocks shown in
Figs.~\ref{fig:result_1} and \ref{fig:result_2}, many genomes seem to
have a significant deviation from the predictions of a strict clock.
Relatedness between genomes will make the observed number of mutations
correlated and not truly independent, rendering statistics such as
$R^2$ and p-values biased and less reliable.  More specifically, a
large fraction of samples collected in 2022 have almost identical
number of synonymous mutations. Any fitting method that passes through
the center of these samples will have a high $R^2$.

In order to minimize the deviation from predictions of a strict
molecular clock, we considered the \textit{principal lineage} defined
by the longest branch length identified by the phylogenetic method
described in the next section, shown in Appendix {\bf Lineage
  analysis} and Fig.~\ref{fig:result_3}.  For the 22 genomes sampled
along this the principal lineage, the linear fit between the number of
synonymous mutations to collection date yields a slope of $5.17$
synonymous mutations per year. As shown in
Fig.~\ref{fig:result_clock}, this fit is quite good with $R^2 =
0.997$.

\paragraph{Weighted least square fitting is preferred for dating the MRCA.}
When the strict molecular clock assumption of constant mutation rates
is valid, the precise number of synonymous mutations with respect to
time is given by a constant-rate Poisson process.  After a given time
$t$, both the mean and the variance in the number of synonymous
mutations is proportional to $t$.

To account for the change in variance, we use a weighted least-squares
method to fit the data. The weights are inversely proportional to the
variance in the number of mutations, and thus inversely proportional to
time $t$. To convert collection dates to time $t$, we need to have a
primary estimate of the emergence time of the MRCA. This is obtained
by ordinary least-squares fitting of the data and is set to
2016-10-01.  However, due to sampling bias, the variance of the number
of mutations for the whole dataset is not proportional to time $t$, so we
restricted our analysis to the genomes within the principal
lineage of MPXV.

The linear relationship between the mean number of synonymous
mutations and time suggests that no major change in the mutation rate
occurred along the principal lineage, including during the unsampled
period between 2019 and 2021.  A few genomes collected in 2022 have a
different number of synonymous mutations. This deviation may be due to
cross-lineage variations. More observations are needed to test whether
the mutation rate has changed since the 2022 global outbreak.

We then extrapolated the linear fit obtained via weighted
least-squares and found a 95\% confidence interval of the time of
emergence of the ancestor $t^{({\rm ancestor})}\in \text{[2016-10-12,
  2017-05-05]}$. The main source of uncertainty is the uncertainty in
the weighted least-squares fit, where the weighting accounts for
possible correlations between different samples.  This estimate is
close to that of a previously published estimate \cite{Rambaut2022c}.



\section*{Discussion and Conclusions}
We developed a simple molecular clock-based method for
analyzing relative mutation rates that is applicable to a broad class
of DNA evolution models.  Molecular clock theory assumes constant a
evolutionary rate across lineages and generations. However, this
assumption is often violated due to differences in environment and
evolutionary mechanisms. Using a less restrictive assumption of
constant \textit{ratios} of mutation rates, we developed a method to
recalibrate a molecular clock using, instead of time, another
molecular clock. If the \textit{relative} rates of different types of
mutations are constant, different lineages from the same ancestor will
have the same \textit{relative} numbers of these different types of
mutations. When we applied our analysis to monkeypox virus (MPXV), we
find that there are two distinct groups of MPXV genomes, those
collected before 2016 and those after. Each group adheres to the
recalibrated molecular clock prediction, with unique relative mutation
rates, indicating different evolutionary pressures potentially caused
by variations in animal and human hosts. The post-2016 group of
sequences \textit{share} biased hypermutations of the pattern TC
$\to$ TT, which is characteristic of APOBEC3-induced
editing. Imposing an additional assumption of independence between
sites, we are able to statistically characterize the initial common
ancestor of lineages appearing after 2016.
%
%
Further invoking an infinite sites assumption mathematically
simplifies the relation between different types of mutations and the
overall analysis. These assumptions allowed us to predict a mean
number of mutations that increases linearly with time (Eqs.~\ref{k1},
\ref{multi-asymp}, and \ref{eq:asymp}), as well as a linear
relationship between the mutations of a specific type to the total
number of mutations (Eqs. \ref{single-linear} and \ref{multi-linear}),
respectively.

Applying our analysis to MPXV samples shows that they evolved in a way
that is quantitatively consistent with our assumptions and that there
are two distinct epochs of mutations characterized by differences in
the rates of the specific dinucleotide mutation TC $\to$ TT, typical
of APOBEC3 editing. \added{We ruled out several
  alternative possibilities for the high APOBEC3-relevant mutation
  frequencies such as selection or drift in animal hosts, or
  differences in the number of available mutation sites.  Dynamic
  biases in mutation rates driven by physical or chemical mutagens are
  also unlikely since the pre-2016 group of MPXV genomes represent a
  baseline that incorporates the effects of mutagens other than
  human-specific enzymes.}

\added{Additionally, our analysis not only identified
  the statistical departure, but also showed that the relative
  mutation rate of TC $\to$ TT is preserved in the pre-2016 and the
  post-2016 sequences, respectively, as shown in
  Fig.~\ref{fig:result_1}.  The tight linear proportionality observed
  is also worth reporting and further investigation. It may reflect a
  relatively homogeneous environment for different individuals in the
  host populations.} \added{It can also allow one
    to discern subtle changes in relative mutation rates.  In
    particular, our analysis suggests that the KJ642617 sequence is
    not a direct ancestor of the post-2016 group, despite it being
    very close.  Similarly, the MK783029 sequence in the post-2016
    group is very likely to have undergone a few APOBEC3-editing
    events since the MRCA of the post-2016 group.}

\begin{figure}[htbp]
  \centering 
  \includegraphics[width=\textwidth]{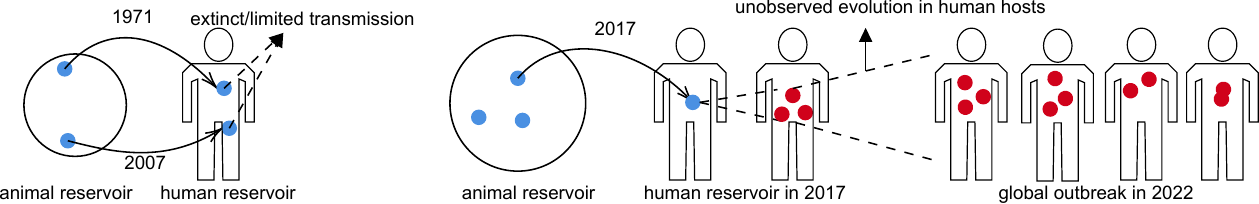}
  \caption{Different evolutionary environments shape the shared
    features of pre- and post-2016 groups of MPXV genomes. Pre-2016
    cross-species transmission did not persist in human hosts. The
    blue dots represent the shared features due to evolution in the
    animal reservoir. The red dots represent the shared features due
    to evolution in the human hosts. We hypothesize an unobserved
    stage of continuous and persistent evolution in the human hosts
    between the 2017 cross-species transmission and the 2022 global
    outbreak.}
  \label{fig:summ}
\end{figure}

We conclude that the most likely scenario is that pre-2016, MPXV
genomes evolved in the animal host. The post-2016 group of MPXV
genomes has undergone persistent and continuous human APOBEC3 editing
after zoonotic transmission circa late 2016/early 2017. This scenario
is implied in Fig.~\ref{fig:summ}.

Unlike applications of DNA evolution models to phylogenetic inference
\cite{Posada2013may}, our method provides a simple and vivid
goodness-of-fit measure by checking whether different genomes share
the same linear relationship between different types of mutations.
While high APOBEC3-relevant mutation rates have been observed and
reported in previous studies, our evolutionary model-based approach
delineates shared features within and between the pre- and post-2016
groups of MPXV genomes. Our model also allows us to identify and
characterize the critical ancestor of the post-2016 group, which may
contain information about post-2016 epidemics. Quantification of the
molecular clock of MPXV suggested that they have a constant mutation
rate in the principal lineage of the post-2016 group. This allowed us
to estimate the time of the ancestral zoonotic transmission leading to
the post-2016 MPXV genomes.

We further validated \replaced{strong APOBEC3-editing in post-2016
  genomes}{our model} by comparing the lineage analysis results with
the results from the BLAST Tree algorithm. Both trees show similar
clusters of genomes, but our method provides better resolution of the
subtle phylogenetic relations ($\preceq$) between genomes collected in
2017 and 2018.
  
We have not explored the quantitative interpretations of connections
between the rate of synonymous mutations and the rate of nonsynonymous
mutations; nonetheless, our linear model and analysis can be directly
applied to other viruses experiencing APOBEC3-driven evolution. For
example, the synonymous mutation rate of SARS-Cov-2 was found to be
relatively stable, while its nonsynonymous mutation rate varied over
time \cite{neher_contributions_2022}. There have been some research
into the relationship between the mutation rates of synonymous and
nonsynonymous mutations
\cite{Spielman2016aug,Bloom2014jul,Zhang2005aug} but they have focused
on improving the inference of the phylogenetic tree. Low levels of
nonsynonymous mutations have already been observed in Feline
parvovirus and Canine parvovirus, and are related to strong selection
\cite{Wang2022}.  \deleted{As we have argued, the mutation probability
  of each synonymous mutation per replication event is relatively
  constant over time and across different genomes. Therefore, the
  number of accumulated synonymous mutations is proportional to the
  number of replication events.}

\added{The idea of inferring evolutionary insights from mutation
  ratios, such as the nonsynonymous to synonymous mutation ratio
  (dN/dS), is well established and has been used to infer selective
  pressures on protein-coding
  regions\cite{Spielman2016aug,Bloom2014jul,Zhang2005aug}. Other
  methods have been developed to identify site-specific mutation rate
  shifts\cite{Duchemin2022dec,Murrell2012may,Yang1994jul}. However,
  these methods often involve phylogenetic tree reconstruction using
  computationally intensive approaches like maximum likelihood or
  Bayesian inference, and they lack an intuitive goodness-of-fit
  measure. In contrast, our method bypasses the need for phylogenetic
  trees, utilizes only sequence alignments against a common reference,
  and offers an intuitive way to measure fit to data.} We can also
evaluate the probability of nonsynonymous mutations per replication,
but this can depend on selection since past nonsynonymous mutations
would have likely changed the fitness of the genome.
A sequence with more potentially beneficial mutations is more likely
to exhibit a higher effective rate of nonsynonymous mutations, while
excess deleterious mutations lead to slower nonsynonymous mutation.
While the basic mutation rate is not directly affected by individual
mutations, beneficial mutations are more likely to survive and spread,
whereas deleterious ones tend to be eliminated. The effective mutation
rate, influenced by selection, provides insight into the evolutionary
landscape.
Therefore, APOBEC3-driven (nonsynonymous) mutations may disrupt the
dynamic mutation-selection balance, which may profoundly influence the
evolution of molecular phenotypes under stabilizing selection
\cite{Rouzine2008feb,Goyal2012aug,Nourmohammad2013jan}. Finding
beneficial mutations is mathematically similar to the problem of a
high-dimensional random walk searching for target sites.  The interplay
between target (beneficial mutations) search and APOBEC3 editing may
be an interesting direction of for future analysis. Combining these
features with previously developed virus dynamics models could yield a
more realistic picture of virus evolution in human
\cite{Lord2021,Kreger2021}.

\section*{Acknowledgements}
TC acknowledges support from the NIH through grant R01HL146552.

\section*{Author Contributions}
XL conceived and constructed the evolution model and performed the
statistical analysis. SH and OY conceived the initial problem and
constructed the virus phylogeny.  XL, SH, and TC wrote the
manuscript. TC and OY assisted in editing the manuscript.

\section*{Disclosure and competing interests}
The authors have no competing interests.


\bibliography{reference}

\section*{Data availability}
The computer code produced in this study are available at:
\begin{itemize}
    \item Analysis computer scripts: Github (\url{github.com/hsianktin/monkeypox}).
\end{itemize}
This study includes no data deposited in external repositories.

\newpage

\appendix
\section*{Supporting Information}
\renewcommand{\thefigure}{S\arabic{figure}}
\renewcommand{\thetable}{S\arabic{table}}
\setcounter{figure}{0}
\setcounter{table}{0}

\subsection*{Hypothetical scenarios}
In this paper, we proposed an idea of visualizing the evolution of
genomes by plotting the number of a specific type of mutations against
the number of synonymous mutations.  

Although alternative methods of detecting mutation rate changes in the
evolution of genomes exist, including the UNREST method and
root-to-tip regression over the reconstructed phylogenetic trees, as
utilized by O'Toole and Rambaut
\cite{Rambaut2022a,Rambaut2022b,Rambaut2022c}, our method has several
advantages. First, it provides a visual presentation for
goodness-of-fit whereas phylogenetic tree-based methods often do not
provide an easily interpretable statistic for goodness-of-fit. Our
method not only detects the change of relative mutation rates, but
also reveals a preserved set of relative mutation rates within the
pre-2016 and post-2016 groups.

Variations in the number of APOBEC3-relevant mutations is smaller than
those theoretically predicted, suggesting additional mechanisms or
processes that may have contributed to controlling the ratios of
mutation rates. This observation invites further study.
A closely followed linear relationship allows us to extrapolate the
linear fit and detect small deviations from the linear relationship.
In the case of monkeypox virus, the grouping is primarily based on the
collection date. However, the lack of the direct ancestor sequence
makes it hard to determine whether or not the MK783029 sequence has
already undergone APOBEC3-editing. As shown in
Fig.~\ref{fig:result_2}(a), a simple calculation of mutation rates for
the MK783029 sequence would suggest that it has similar relative
mutation rates as the pre-2016 group since it just entered human hosts
and had not undergone much APOBEC3-editing. By using MK783029 as the
reference genome, and using linear fitting to extrapolate both the
relative mutation rates and the variation of the rates, we can detect
subtle changes in relative mutation rates.

\begin{figure}[h!]
  \centering 
  \includegraphics[width=0.8\textwidth]{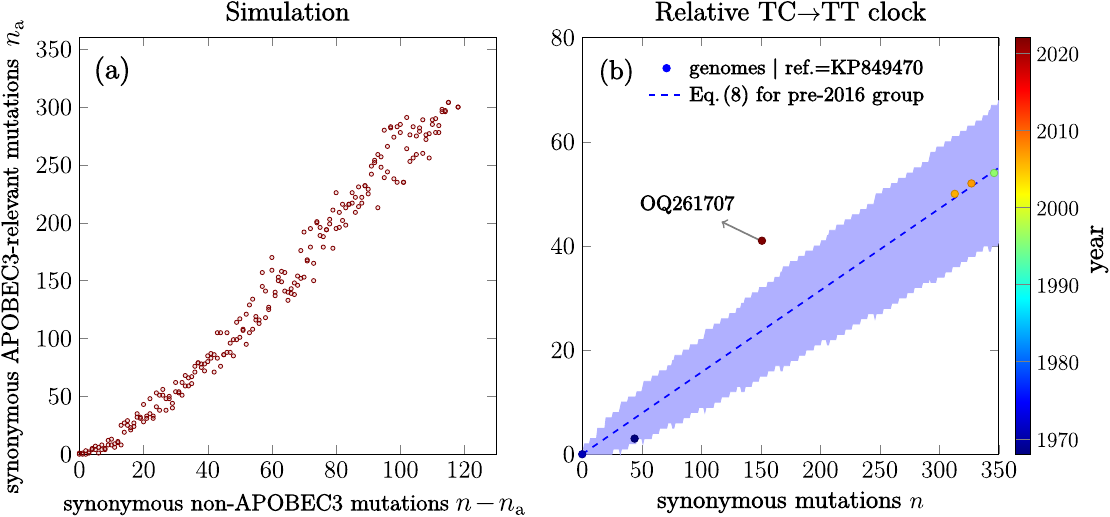}
  \caption{\added{Hypothetical scenarios. (a) A Markov chain simulation
    with $N=10^6, N_{\rm +a}=10^{4}$. The mutation rate for
    other synonymous mutations is $10^{-7}$ per unit time per site,
    while the APOBEC3 mutation rate is set to $10^{-5} (1 +
    t/200)$. Simulations are stopped at $t=1000$.  (b) A real-world
    example with limited samples where only one sequence, OQ261707,
    from the post-2016 group is available and the sequence KJ642617 is
    not available in the pre-2016 group. The shaded region represents
    the 95\% prediction interval of the samples following the same
    linear relationship as the pre-2016 group. The sequence OQ261707
    lies outside the 95\% prediction interval suggesting a significant
    change of relative mutation rates.}}
  \label{fig:resp_2}
\end{figure}

We now discuss two hypothetical scenarios under which our approach may be
particularly useful. The first example is a simulated scenario wherein
the activity of APOBEC3-editing increases linearly over time. This
leads to a quadratic relationship between the number of
APOBEC3-relevant mutations ($n_{\rm a}$) and the number of synonymous
non-APOBEC3-relevant mutations ($n - n_{\rm a}$), as shown in
Fig.~\ref{fig:resp_2}(a).

Our second example is rooted in real-world data derived from currently
available genome sequences. Earlier evidence by O'Toole \textit{et
  al}. of APOBEC3-editing in the post-2016 genomes hinges on the
presence of multiple post-2016 genomes sampled at different time
points and a genome (KJ642617) that is closely related to the
MRCA. These contributed to a root-to-tip regression estimate of the
APOBEC3 editing rate. Now, consider a hypothetical scenario where the
genome KJ642617 is not available, and only a single genome from the
post-2016 group, namely OQ261707 sampled in 2022, is accessible. In
this situation, a comparison of OQ261707 to any other genome from the
pre-2016 group does not reveal a dominance of APOBEC3-signature
mutations over other synonymous mutations, as depicted in
Fig.~\ref{fig:resp_2}(b). To discern the APOBEC3-editing activity from
natural mutations, reconstructing the phylogenetic tree and ancestral
sequences becomes challenging and introduces further layers of
uncertainty.

However, our method proves effective in this scenario as it
automatically generates the 95\% prediction interval for the number of
APOBEC3-relevant mutations ($n_{\rm a}$) for genome OQ261707 under the
null hypothesis that APOBEC3-editing activity is absent, as shown by
the shaded region in Fig.~\ref{fig:resp_2}(b). The observed number
$n_{\rm a}$ of APOBEC3-relevant mutations considerably exceeds the
95\% prediction interval indicating, even with such limited data, the
significance of the APOBEC3-editing activity.

\paragraph*{S1 Table.}
\label{S1_Table}
Accession numbers and dates used in this study. When the exact date of
collection is not known, we round the date to be beginning of the
known year or month.
\setlength\tabcolsep{30pt}

\begin{filecontents*}{table_s1.tex}
Accession,Date,Clade,Reference
KJ642616,1968-01-01,IIa,\cite{milhaud1969analyse,Nakazawa2015apr}
KJ642617,1971-01-01,IIb,\cite{foster1972human,Nakazawa2015apr}
KP849470,1971-01-01,IIa,\cite{arita1976monkeypox,Nakazawa2015apr}
NC\_003310,1996-01-01,I,\cite{hutin2001outbreak}
JX878410,2006-11-24,I,\cite{kugelman2014genomic}
JX878428,2007-06-30,I,\cite{kugelman2014genomic}
OP535341,2017-10-01,IIb,\cite{Nnaemeka2022unpublished}
OP535340,2017-10-01,IIb,\cite{Nnaemeka2022unpublished}
OP535337,2017-10-01,IIb,\cite{Nnaemeka2022unpublished}
OP535336,2017-10-01,IIb,\cite{Nnaemeka2022unpublished}
OP535335,2017-10-01,IIb,\cite{Nnaemeka2022unpublished}
OP535323,2017-10-01,IIb,\cite{Nnaemeka2022unpublished}
OP535322,2017-10-01,IIb,\cite{Nnaemeka2022unpublished}
MK783032,2017-11-01,IIb,\cite{yinka2019outbreak}
OP535320,2017-11-01,IIb,\cite{Nnaemeka2022unpublished}
MK783028,2017-11-09,IIb,\cite{yinka2019outbreak}
MK783031,2017-11-09,IIb,\cite{yinka2019outbreak}
MK783030,2017-11-30,IIb,\cite{yinka2019outbreak}
MK783029,2017-12-06,IIb,\cite{yinka2019outbreak}
OP535333,2018-01-01,IIb,\cite{Nnaemeka2022unpublished}
OP535324,2018-02-01,IIb,\cite{Nnaemeka2022unpublished}
OP535312,2018-02-01,IIb,\cite{Nnaemeka2022unpublished}
OP535329,2018-03-01,IIb,\cite{Nnaemeka2022unpublished}
OP535328,2018-03-01,IIb,\cite{Nnaemeka2022unpublished}
OP535327,2018-03-01,IIb,\cite{Nnaemeka2022unpublished}
OP535325,2018-06-01,IIb,\cite{Nnaemeka2022unpublished}
NC\_063383,2018-08-01,IIb,\cite{mauldin2020exportation}
MT903341,2018-08-14,IIb,\cite{mauldin2020exportation}
MN648051,2018-10-04,IIb,\cite{cohen2020identification}
MT903344,2018-10-09,IIb,\cite{mauldin2020exportation}
MT903345,2018-10-09,IIb,\cite{mauldin2020exportation}
MT903343,2018-10-17,IIb,\cite{mauldin2020exportation}
MT903342,2019-04-30,IIb,\cite{mauldin2020exportation}
ON676708,2021-11-01,IIb,\cite{gigante2022multiple}
ON563414,2022-05-01,IIb,\cite{gigante2022multiple}
ON694329,2022-05-01,IIb,\cite{brinkmann2022possible}
OP120937,2022-05-01,IIb,\cite{AjaMacaya2022medRxiv}
ON649713,2022-05-19,IIb,\cite{Isidro2022jun}
ON843173,2022-05-27,IIb,\cite{Isidro2022jun}
ON843172,2022-05-27,IIb,\cite{Isidro2022jun}
ON843174,2022-05-30,IIb,\cite{Isidro2022jun}
OP225968,2022-06-01,IIb,\cite{gigante2022multiple}
ON813267,2022-06-01,IIb,\cite{brinkmann2022possible}
ON813266,2022-06-01,IIb,\cite{brinkmann2022possible}
ON813261,2022-06-01,IIb,\cite{brinkmann2022possible}
ON813255,2022-06-01,IIb,\cite{brinkmann2022possible}
ON813251,2022-06-01,IIb,\cite{brinkmann2022possible}
OP764628,2022-06-01,IIb,\cite{brinkmann2022possible}
OP604533,2022-06-01,IIb,\cite{gigante2022multiple}
OP604532,2022-06-01,IIb,\cite{gigante2022multiple}
OP604530,2022-06-01,IIb,\cite{gigante2022multiple}
OP604529,2022-06-01,IIb,\cite{gigante2022multiple}
OP604528,2022-06-01,IIb,\cite{gigante2022multiple}
OP604527,2022-06-01,IIb,\cite{gigante2022multiple}
OP604522,2022-06-01,IIb,\cite{gigante2022multiple}
OP837353,2022-06-01,IIb,\cite{brinkmann2022possible}
OP837352,2022-06-01,IIb,\cite{brinkmann2022possible}
OP837351,2022-06-01,IIb,\cite{brinkmann2022possible}
OP837350,2022-06-01,IIb,\cite{brinkmann2022possible}
OP837349,2022-06-01,IIb,\cite{brinkmann2022possible}
OP837347,2022-06-01,IIb,\cite{brinkmann2022possible}
OP837346,2022-06-01,IIb,\cite{brinkmann2022possible}
OP837345,2022-06-01,IIb,\cite{brinkmann2022possible}
OP837344,2022-06-01,IIb,\cite{brinkmann2022possible}
OP837343,2022-06-01,IIb,\cite{brinkmann2022possible}
OP837342,2022-06-01,IIb,\cite{brinkmann2022possible}
OP837341,2022-06-01,IIb,\cite{brinkmann2022possible}
OP837340,2022-06-01,IIb,\cite{brinkmann2022possible}
OP837339,2022-06-01,IIb,\cite{brinkmann2022possible}
OP837338,2022-06-01,IIb,\cite{brinkmann2022possible}
OP837337,2022-06-01,IIb,\cite{brinkmann2022possible}
OP837336,2022-06-01,IIb,\cite{brinkmann2022possible}
OP837335,2022-06-01,IIb,\cite{brinkmann2022possible}
OP837334,2022-06-01,IIb,\cite{brinkmann2022possible}
OQ099608,2022-06-01,IIb,\cite{gigante2022multiple}
OP120938,2022-06-01,IIb,\cite{AjaMacaya2022medRxiv}
ON843176,2022-06-02,IIb,\cite{Isidro2022jun}
OQ249660,2022-06-08,IIb,\cite{Baliere2023unpublished}
OP536686,2022-06-09,IIb,\cite{Gagnon2022unpublished}
OP536688,2022-06-12,IIb,\cite{Gagnon2022unpublished}
OQ249661,2022-06-14,IIb,\cite{Baliere2023unpublished}
OP536697,2022-06-17,IIb,\cite{Gagnon2022unpublished}
OP536702,2022-06-21,IIb,\cite{Gagnon2022unpublished}
OP536704,2022-06-22,IIb,\cite{Gagnon2022unpublished}
OP536709,2022-06-23,IIb,\cite{Gagnon2022unpublished}
OP536712,2022-06-24,IIb,\cite{Gagnon2022unpublished}
OP536711,2022-06-24,IIb,\cite{Gagnon2022unpublished}
OP536713,2022-06-25,IIb,\cite{Gagnon2022unpublished}
OP536731,2022-06-30,IIb,\cite{Gagnon2022unpublished}
OP123044,2022-07-01,IIb,\cite{sereewit2022orf}
OP123045,2022-07-01,IIb,\cite{sereewit2022orf}
OP123041,2022-07-01,IIb,\cite{sereewit2022orf}
OP123040,2022-07-01,IIb,\cite{sereewit2022orf}
OP536738,2022-07-01,IIb,\cite{Gagnon2022unpublished}
OP536737,2022-07-01,IIb,\cite{Gagnon2022unpublished}
OP536732,2022-07-01,IIb,\cite{Gagnon2022unpublished}
OP743951,2022-07-01,IIb,\cite{sereewit2022orf}
OP434519,2022-07-01,IIb,\cite{sereewit2022orf}
OP392531,2022-07-01,IIb,\cite{sereewit2022orf}
OP604531,2022-07-01,IIb,\cite{gigante2022multiple}
OP604526,2022-07-01,IIb,\cite{gigante2022multiple}
OP604525,2022-07-01,IIb,\cite{gigante2022multiple}
OP604524,2022-07-01,IIb,\cite{gigante2022multiple}
OP604523,2022-07-01,IIb,\cite{gigante2022multiple}
OP604521,2022-07-01,IIb,\cite{gigante2022multiple}
OP604520,2022-07-01,IIb,\cite{gigante2022multiple}
OP604519,2022-07-01,IIb,\cite{gigante2022multiple}
OP881956,2022-07-01,IIb,\cite{gigante2022multiple}
OP881955,2022-07-01,IIb,\cite{gigante2022multiple}
OP881954,2022-07-01,IIb,\cite{gigante2022multiple}
OQ054242,2022-07-01,IIb,\cite{gigante2022multiple}
OQ099609,2022-07-01,IIb,\cite{gigante2022multiple}
OQ099607,2022-07-01,IIb,\cite{gigante2022multiple}
OQ099606,2022-07-01,IIb,\cite{gigante2022multiple}
OQ099605,2022-07-01,IIb,\cite{gigante2022multiple}
OQ099604,2022-07-01,IIb,\cite{gigante2022multiple}
OP536742,2022-07-02,IIb,\cite{Gagnon2022unpublished}
OP536740,2022-07-02,IIb,\cite{Gagnon2022unpublished}
OP536745,2022-07-06,IIb,\cite{Gagnon2022unpublished}
OP536718,2022-07-06,IIb,\cite{Gagnon2022unpublished}
OP879722,2022-07-26,IIb,\cite{Palmateer2022unpublished}
OP879723,2022-07-27,IIb,\cite{Palmateer2022unpublished}
OP257247,2022-08-01,IIb,\cite{sereewit2022orf}
OP257243,2022-08-01,IIb,\cite{sereewit2022orf}
OP215267,2022-08-01,IIb,\cite{brinkmann2022possible}
OP743962,2022-08-01,IIb,\cite{sereewit2022orf}
OP743961,2022-08-01,IIb,\cite{sereewit2022orf}
OP743960,2022-08-01,IIb,\cite{sereewit2022orf}
OP743959,2022-08-01,IIb,\cite{sereewit2022orf}
OP743958,2022-08-01,IIb,\cite{sereewit2022orf}
OP743957,2022-08-01,IIb,\cite{sereewit2022orf}
OP743956,2022-08-01,IIb,\cite{sereewit2022orf}
OP743955,2022-08-01,IIb,\cite{sereewit2022orf}
OP743954,2022-08-01,IIb,\cite{sereewit2022orf}
OP743953,2022-08-01,IIb,\cite{sereewit2022orf}
OP743952,2022-08-01,IIb,\cite{sereewit2022orf}
OP392553,2022-08-01,IIb,\cite{sereewit2022orf}
OP392552,2022-08-01,IIb,\cite{sereewit2022orf}
OP392551,2022-08-01,IIb,\cite{sereewit2022orf}
OP392549,2022-08-01,IIb,\cite{sereewit2022orf}
OP392548,2022-08-01,IIb,\cite{sereewit2022orf}
OP392547,2022-08-01,IIb,\cite{sereewit2022orf}
OP392546,2022-08-01,IIb,\cite{sereewit2022orf}
OP392545,2022-08-01,IIb,\cite{sereewit2022orf}
OP392544,2022-08-01,IIb,\cite{sereewit2022orf}
OP392543,2022-08-01,IIb,\cite{sereewit2022orf}
OP392542,2022-08-01,IIb,\cite{sereewit2022orf}
OP392541,2022-08-01,IIb,\cite{sereewit2022orf}
OP392540,2022-08-01,IIb,\cite{sereewit2022orf}
OP392539,2022-08-01,IIb,\cite{sereewit2022orf}
OP392538,2022-08-01,IIb,\cite{sereewit2022orf}
OP392536,2022-08-01,IIb,\cite{sereewit2022orf}
OP392535,2022-08-01,IIb,\cite{sereewit2022orf}
OP392534,2022-08-01,IIb,\cite{sereewit2022orf}
OP392533,2022-08-01,IIb,\cite{sereewit2022orf}
OP392532,2022-08-01,IIb,\cite{sereewit2022orf}
OP881952,2022-08-01,IIb,\cite{gigante2022multiple}
OP881951,2022-08-01,IIb,\cite{gigante2022multiple}
OP881950,2022-08-01,IIb,\cite{gigante2022multiple}
OP881948,2022-08-01,IIb,\cite{gigante2022multiple}
OP881941,2022-08-01,IIb,\cite{gigante2022multiple}
OQ054241,2022-08-01,IIb,\cite{gigante2022multiple}
OQ054235,2022-08-01,IIb,\cite{gigante2022multiple}
OQ054229,2022-08-01,IIb,\cite{gigante2022multiple}
OP820455,2022-08-08,IIb,\cite{Barbian2022unpublished}
OQ121956,2022-08-28,IIb,\cite{Barbian2022unpublished}
OP743993,2022-09-01,IIb,\cite{sereewit2022orf}
OP743992,2022-09-01,IIb,\cite{sereewit2022orf}
OP743991,2022-09-01,IIb,\cite{sereewit2022orf}
OP743990,2022-09-01,IIb,\cite{sereewit2022orf}
OP743989,2022-09-01,IIb,\cite{sereewit2022orf}
OP743988,2022-09-01,IIb,\cite{sereewit2022orf}
OP743987,2022-09-01,IIb,\cite{sereewit2022orf}
OP743986,2022-09-01,IIb,\cite{sereewit2022orf}
OP743985,2022-09-01,IIb,\cite{sereewit2022orf}
OP743984,2022-09-01,IIb,\cite{sereewit2022orf}
OP743983,2022-09-01,IIb,\cite{sereewit2022orf}
OP743982,2022-09-01,IIb,\cite{sereewit2022orf}
OP743981,2022-09-01,IIb,\cite{sereewit2022orf}
OP743980,2022-09-01,IIb,\cite{sereewit2022orf}
OP743979,2022-09-01,IIb,\cite{sereewit2022orf}
OP743978,2022-09-01,IIb,\cite{sereewit2022orf}
OP743976,2022-09-01,IIb,\cite{sereewit2022orf}
OP743975,2022-09-01,IIb,\cite{sereewit2022orf}
OP743974,2022-09-01,IIb,\cite{sereewit2022orf}
OP743973,2022-09-01,IIb,\cite{sereewit2022orf}
OP743972,2022-09-01,IIb,\cite{sereewit2022orf}
OP743971,2022-09-01,IIb,\cite{sereewit2022orf}
OP743970,2022-09-01,IIb,\cite{sereewit2022orf}
OP743969,2022-09-01,IIb,\cite{sereewit2022orf}
OP743968,2022-09-01,IIb,\cite{sereewit2022orf}
OP743967,2022-09-01,IIb,\cite{sereewit2022orf}
OP743966,2022-09-01,IIb,\cite{sereewit2022orf}
OP743965,2022-09-01,IIb,\cite{sereewit2022orf}
OP743964,2022-09-01,IIb,\cite{sereewit2022orf}
OP715789,2022-09-01,IIb,\cite{sereewit2022orf}
OP715788,2022-09-01,IIb,\cite{sereewit2022orf}
OP881953,2022-09-01,IIb,\cite{gigante2022multiple}
OP881949,2022-09-01,IIb,\cite{gigante2022multiple}
OP881947,2022-09-01,IIb,\cite{gigante2022multiple}
OP881946,2022-09-01,IIb,\cite{gigante2022multiple}
OP881945,2022-09-01,IIb,\cite{gigante2022multiple}
OP881944,2022-09-01,IIb,\cite{gigante2022multiple}
OP881943,2022-09-01,IIb,\cite{gigante2022multiple}
OP881942,2022-09-01,IIb,\cite{gigante2022multiple}
OQ054247,2022-09-01,IIb,\cite{gigante2022multiple}
OQ054246,2022-09-01,IIb,\cite{gigante2022multiple}
OQ054245,2022-09-01,IIb,\cite{gigante2022multiple}
OQ054244,2022-09-01,IIb,\cite{gigante2022multiple}
OQ054243,2022-09-01,IIb,\cite{gigante2022multiple}
OQ054240,2022-09-01,IIb,\cite{gigante2022multiple}
OQ054239,2022-09-01,IIb,\cite{gigante2022multiple}
OQ054238,2022-09-01,IIb,\cite{gigante2022multiple}
OQ054237,2022-09-01,IIb,\cite{gigante2022multiple}
OQ054236,2022-09-01,IIb,\cite{gigante2022multiple}
OQ054234,2022-09-01,IIb,\cite{gigante2022multiple}
OQ054233,2022-09-01,IIb,\cite{gigante2022multiple}
OQ054232,2022-09-01,IIb,\cite{gigante2022multiple}
OQ054231,2022-09-01,IIb,\cite{gigante2022multiple}
OQ054230,2022-09-01,IIb,\cite{gigante2022multiple}
OQ054228,2022-09-01,IIb,\cite{gigante2022multiple}
OQ054227,2022-09-01,IIb,\cite{gigante2022multiple}
OQ054226,2022-09-01,IIb,\cite{gigante2022multiple}
OQ054225,2022-09-01,IIb,\cite{gigante2022multiple}
OQ054224,2022-09-01,IIb,\cite{gigante2022multiple}
OQ054223,2022-09-01,IIb,\cite{gigante2022multiple}
OQ054222,2022-09-01,IIb,\cite{gigante2022multiple}
OQ054221,2022-09-01,IIb,\cite{gigante2022multiple}
OQ054220,2022-09-01,IIb,\cite{gigante2022multiple}
OQ054219,2022-09-01,IIb,\cite{gigante2022multiple}
OQ054218,2022-09-01,IIb,\cite{gigante2022multiple}
OP820454,2022-09-07,IIb,\cite{Barbian2022unpublished}
OP820456,2022-09-08,IIb,\cite{Barbian2022unpublished}
OQ121962,2022-09-25,IIb,\cite{Barbian2022unpublished}
OP764629,2022-10-01,IIb,\cite{brinkmann2022possible}
OP837354,2022-10-01,IIb,\cite{brinkmann2022possible}
OQ261707,2022-10-14,IIb,\cite{Betancur2023unpublished}
\end{filecontents*}
\pgfplotstabletypeset[
		header=true,
		col sep=comma,
        columns = {Accession, Date, Clade, Reference},
        columns/Accession/.style={string type},
        columns/Date/.style={string type},
        columns/Clade/.style={string type},
        columns/Reference/.style={string type},
        every first row/.append style={
          before row={
     \caption{Accession numbers and dates used in this study.}\label{table1}
            \\ \toprule
            {Accession} & {Date} & {Clade} & {Reference} \\ \midrule
            \endfirsthead
          },
        },
		every head row/.style={
		  before row={
       \caption{Accession numbers and dates used in this study.}\\ \toprule},
		  after row={
			  \midrule\endhead}
			  },
		every last row/.style={after row={
			\bottomrule
			}},]{table_s1.tex}

\subsection*{Lineage analysis}
\begin{figure}
  \begin{center}
      \includegraphics[width=2.7in]{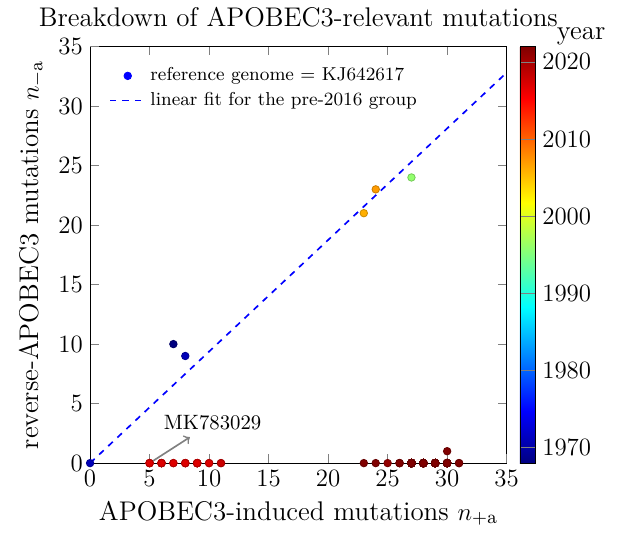}
  \end{center}
      \caption{The number of observed reverse-APOBEC3 mutations
        $n_{\rm -a}$ plotted against the number of observed
        APOBEC3-induced mutations $n_{\rm +a}$, using the genome
        KJ642617 as a reference.}
      \label{fig:supp_2}
\end{figure}

APOBEC3-editing is asymmetric with $v_{\rm +a}\gg v_{{\rm -a}}$. As
shown in Fig.~\ref{fig:supp_2}, few reverse-APOBEC3 synonymous
mutations were observed in the post-2016 group compared to genome
KJ642617.  According to Eq.~\eqref{multi-asymp}, since $v_{\rm +a}\gg
v_{{\rm -a}}$,

\begin{equation}
    \begin{aligned}
        n_{\rm +a}^{(i \mid j)} &\approx N_{{\rm +a}} v_{{\rm +a}}\big(t^{(i)}-t^{(ij)}\big)\\
        n_{\rm -a}^{(i \mid j)} &\approx N_{{\rm +a}} v_{{\rm +a}}\big(t^{(j)}-t^{(ij)}\big)
        \end{aligned}
        \label{eq:asymp}
\end{equation}
By comparing any two genomes from the post-2016 group, the values of
$n_{\rm +a}^{(i\mid j)}$ and $n_{\rm -a}^{(i\mid j)}$ provide
information about the relation among $t^{(i)}$, $t^{(j)}$, and
$t^{(ij)}$. If $n_{\rm +a}^{(i\mid j)}=0, n_{\rm +a}^{(j\mid i)}=0$,
then $t^{(i)}=t^{(j)}=t^{(ij)}$. This suggests that genomes $(i)$ and
$(j)$ are approximately on the same node of the phylogenetic tree. If
$n_{\rm +a}^{(i\mid j)}=0, n_{\rm +a}^{(j\mid i)}>0$, then
$t^{(i)}=t^{(ij)} < t^{(j)}$, suggesting that genome $(i)$ is close to
an ancestor of genome $(j)$.

\begin{figure}[h!]
  \centering 
  \includegraphics[width=\textwidth]{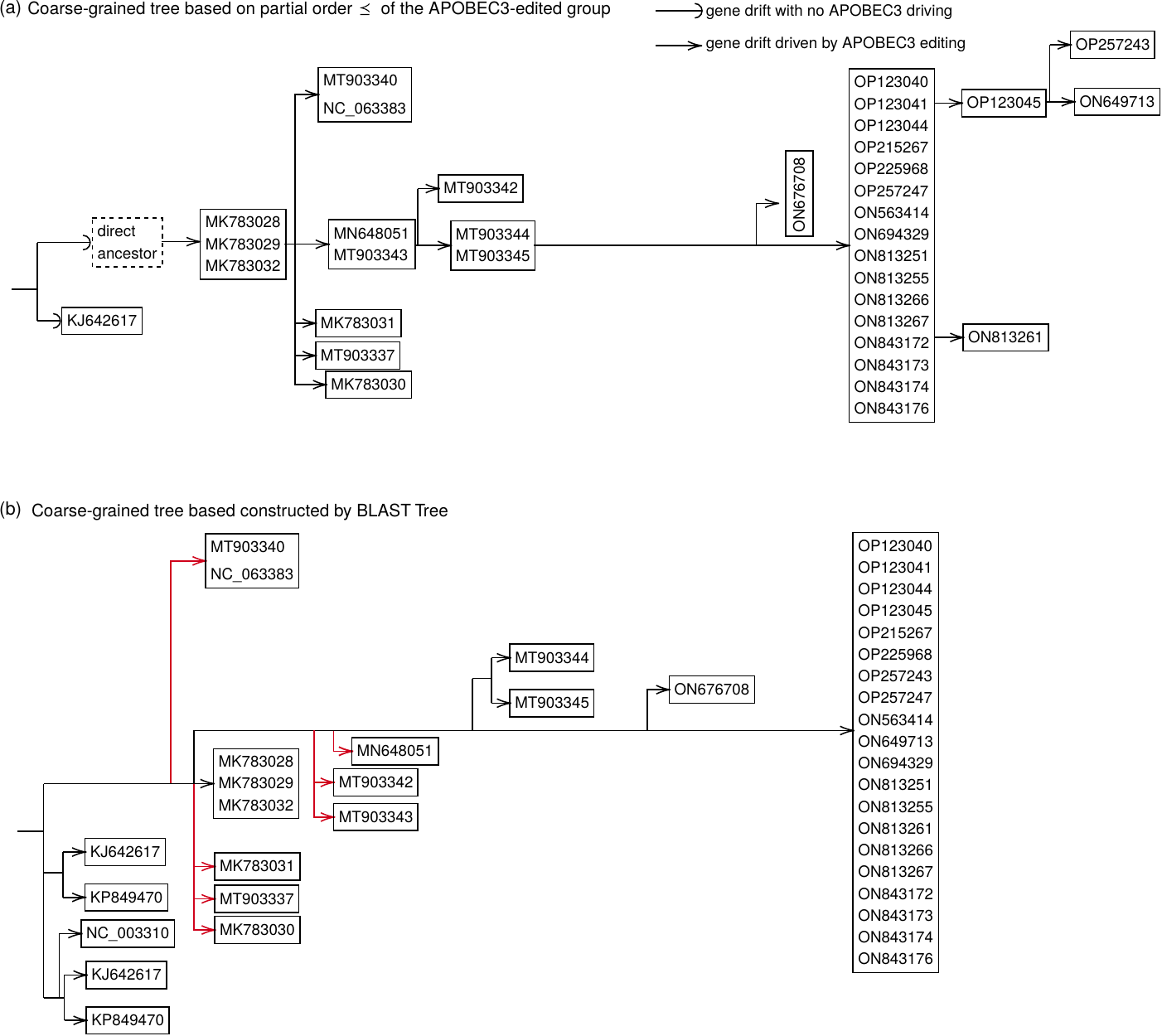}
  \caption{(a) The phylogenetic tree of the post-2016 group and the
    genome KP849470. The tree was constructed by the partial order
    defined by the APOBEC3-induced mutations.  Genomes listed in the
    same box are equivalent in terms of their synonymous
    APOBEC3-induced mutations. Lines with an arrow represent evolution
    within the post-2016 group. Lines without arrows represent neutral
    evolution free from APOBEC3-editing.  The dashed box represents
    the hypothetical direct ancestor of the post-2016 group. (b) The
    phylogenetic tree computed by the BLAST Tree algorithm based on
    pairwise alignments of the 40 MPXV sequences listed in Table
    \ref{table1}.  Similar genomes in (b) are clustered into a
    dashed-line box for simplicity and clarity. Both trees clustered
    similar genomes correctly. However, the ancestral relationship of
    the post-2016 group is not preserved by the BLAST Tree algorithm,
    as the red branches in (b) indicate.}
  \label{fig:result_3}
\end{figure}

We use these comparisons to infer a phylogenetic tree.  By comparing
all genomes in the post-2016 group, we first identify subgroups of
genomes which contain the same APOBEC3-induced mutations. In our
method, genomes within each of these subgroups are indistinguishable
from each other and are thus lumped into the same box in
Fig.~\ref{fig:result_3}(a).  If all APOBEC3-induced mutations in
genome $(i)$ appear in genome $(j)$, then $(i)$ is likely an ancestor
of $(j)$ and $(j)$ inherited the APOBEC3-induced mutations. We denote
this \textit{ancestral relationship} as $i \preceq j$. This
relationship naturally extends to groups of genomes with the same
APOBEC3-induced mutations.

A phylogenetic tree can be constructed by merging all the maximal
chains with respect to the ancestral relationship $\preceq$ into a
single tree. A maximal chain is a chain that cannot be a subsequence
of a longer chain and represents a complete path of evolution in the
observed genomes. For example, suppose that there are 5 groups of
indistinguishable genomes $A, B, C, D, E$ and that $A \preceq B$, $A
\preceq C$, $A \preceq D$, $A \preceq E$, $B \preceq C$, and $B
\preceq D$. Then, all the maximal chains in this set are $A \preceq B
\preceq C$, $A \preceq B \preceq D$, $A \preceq E$. And $A$ is the
common root of the tree, $B$ and $E$ are two direct descendants of
$A$, and $C$ and $D$ are two direct descendants of $B$.

This proposed algorithm to identify maximal chains typically requires
fewer computational resources than other probability-model-based
phylogenetic tree reconstruction methods. The reconstructed tree using
40 representative samples from the 237 genomes shown in
Fig.~\ref{fig:result_3}(a) has similar clusters of genomes as the tree
constructed by the BLAST Tree algorithm \cite{King2010} shown in
Fig.~\ref{fig:result_3}(b), which also relies on pairwise comparisons
of genomes. However, the BLAST algorithm does not correctly resolve
the subtle ancestral relationships between genomes collected in 2017
and 2018, highlighted in red.

\subsection*{Distribution of APOBEC3-induced mutations}
\begin{figure}[h!]
  \begin{center}
      \includegraphics[width=2.7in]{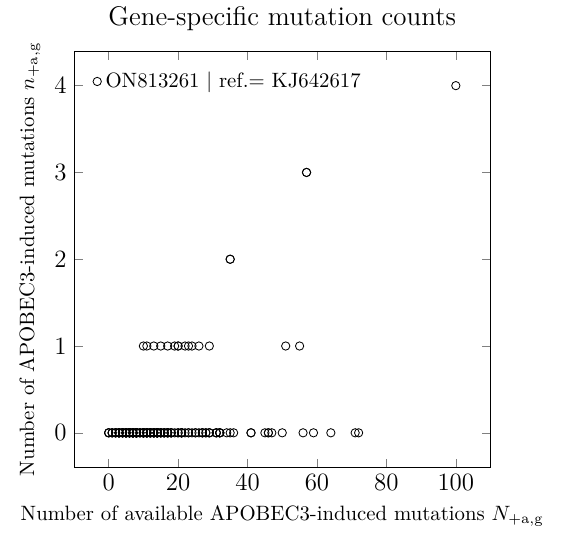}
  \end{center}
     \caption{Using the given genome ON813261, and the reference
        genome KJ642617, we can calculate the number of possible
        APOBEC3-induced mutations and the number of observed mutations
        for \textit{each gene}. Then, each gene is represented by a
        point in the 2D plane. The x-axis represents the number of
        possible APOBEC3-induced mutations, and the y-axis represents
        the number of observed APOBEC3-induced mutations.}
     \label{fig:supp_3}
     \end{figure}

%
Since the APOBEC3 enzyme can diffuse along the DNA strand while
editing, it can potentially generate clusters of APOBEC3-induced
mutations localized to within processive footprints.  On the other
hand, rapid adaptation of MPXV to human hosts may lead to preferred
mutation sites within specific viral genes. To quantify how
APOBEC3-induced mutations are distributed along MPXV genomes, we
counted their number within each gene.  Using KJ642617 as a reference,
the number of possible APOBEC3-induced mutations and the actual number
of APOBEC3-induced mutations are evaluated for each
gene. Fig.~\ref{fig:supp_3} shows a scatter plot of these values for
each gene.

Because the number of observed mutations ($\sim 30$) is much smaller
than the number of genes ($\sim 200$), we cannot draw any statistical
conclusion from this plot. As expected, most genes do not carry any
APOBEC3-induced mutations, and most of the others have only one
mutation.  There are three genes carrying 2, 3, and 4 mutations,
respectively. But these genes do also have more possible mutation
sites than average.  Finally, we also analyzed the correlation between
and within the number of synonymous and nonsynonymous APOBEC3
mutations in each gene from the post-2016 genomes.

\begin{figure}[htb]
      \centering
      \includegraphics[width=5in]{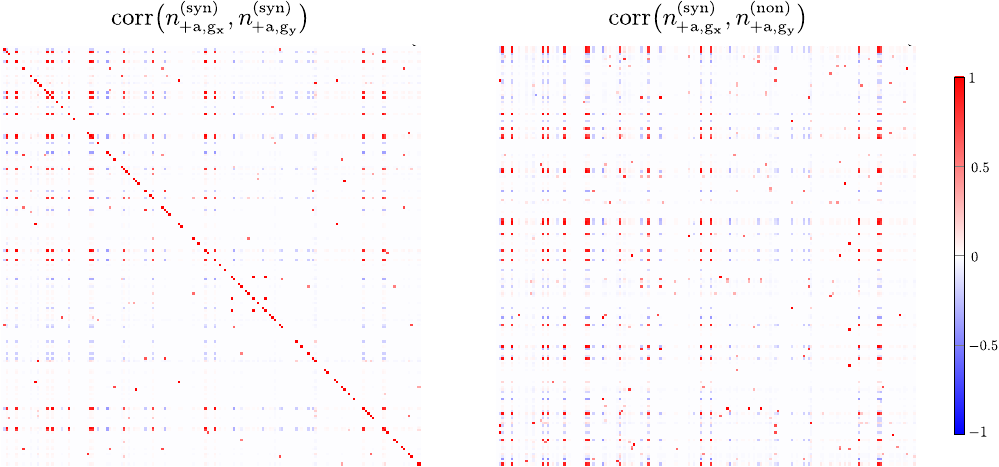}
      \caption{Heatmap of correlation coefficients between
        APOBEC3-induced mutation counts for all pairs of genes using
        reference genome KJ642617 and post-2016 genomes. Each row and
        column represents a gene. Genes are arranged by its location
        along the genome. The left panel shows correlations between
        synonymous APOBEC3-induced mutation counts for each pair of
        genes (corr($n_{\rm {+a, g}_x}^{(\textrm{syn})}$, $n_{\rm {+a,
            g}_y}^{(\textrm{syn})}$)), while the right panel displays
        correlations between synonymous and nonsynonymous mutation
        counts (corr($n_{\rm {+a, g}_x}^{(\textrm{syn})}$, $n_{\rm
          {+a, g}_y}^{(\textrm{non})}$)). In this heatmap, each row
        represents the correlation between nonsynonymous
        APOBEC3-induced mutation counts of a given gene ${\rm g}_y$
        and synonymous APOBEC3-induced mutation counts for different
        genes in order. Similarly, each column represents the
        correlation between synonymous APOBEC3-induced mutation counts
        of a given gene ${\rm g}_x$ and nonsynonymous APOBEC3-induced
        mutation counts for different genes in order.  The pattern of
        each row in the right panel is similar to the pattern of each
        row in the left panel although the patterns of each column in
        two panels are slightly different.}
      \label{fig:supp_4}
\end{figure}

To calculate the correlations between genes, we consider gene-specific
mutation counts as random variables, tally mutations within each gene
for every genome, and use the empirical distribution as the random
variable distribution. For example, suppose genome $(i)$ carries two
mutations in gene 1 and no mutations in gene 2, while genome $(j)$
carries three mutations in gene 1 and one mutation in gene 2.
%
%
We treat the numbers of mutations in gene 1 and gene 2 as two random
variables $n_{\rm g_1}$ and $n_{\rm g_2}$.  Considering only genomes
$(i)$ and $(j)$, the configurations $(n_{\rm g_1}=2, n_{\rm g_2}=0)$
and $(n_{\rm g_1}=3, n_{\rm g_2}=1)$ both arise with probability
$1/2$. In the same way, we obtain the joint distribution of different
random variables corresponding to different genes in the post-2016
genomes and calculate the correlation coefficient between them. During
the enumeration process, we separately consider the number of
synonymous APOBEC3-induced mutations and the number of nonsynonymous
APOBEC3-induced mutations for each gene. Correlations in the number of
mutations across genes are typically low as shown in
Fig.~\ref{fig:supp_4}.

A few highly correlated sites observed are unlikely to arise from
functional relationships among genes. Rather, these correlations arise
from relatedness since most of the genomes collected in 2022 and are
from a single lineage that acquired some mutations between 2018 and
2021.
%
%
By this same relatedness among sample genomes, there is a strong
correlation between the number of synonymous APOBEC3-induced mutations
of the same subset of genes and the number of nonsynonymous
APOBEC3-induced mutations for a different subset of genes, as is shown
in the right panel of Fig.~\ref{fig:supp_4}.  Since the set of genes
exhibiting correlations depends on the mutations that arose between
2018 and 2021, it is not surprising that for nonsynonymous mutations,
there is a different set of strongly correlated genes. Overall, there
is little structure in the sequence-dependence of mutations,
consistent with our independent sites assumption.

\begin{figure}[htbp]
      \centering
\includegraphics[width=5in]{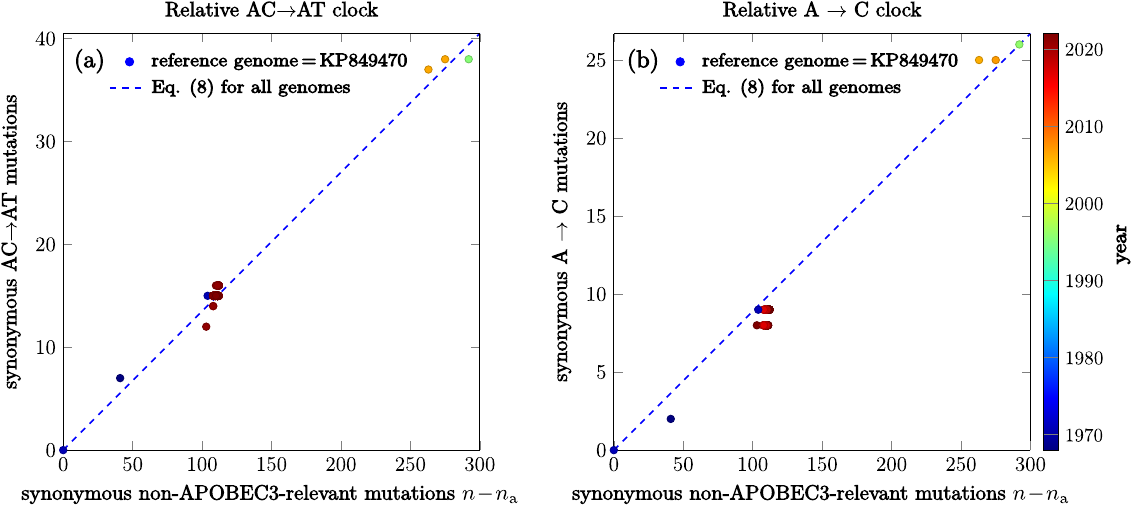}
\caption{\added{Relative molecular clocks for different types of
    mutations with respect to KP849470, instead of time, exhibit
    linearity. (a) The number of $\textrm{AC}\to\textrm{AT}$ mutations
    for each genome, plotted relative to its total number of
    synonymous mutations excluding APOBEC3-relevant mutations. (b) The
    number of $\textrm{A}\to \textrm{C}$ mutations for each genome,
    plotted relative to its total number of synonymous mutations
    excluding APOBEC3-relevant mutations. The y-axis in (a) and (b)
    represent ``other mutations'' as is shown in green in
    Fig.~\ref{fig-schematic}(a).}}
      \label{fig:supp_1}
\end{figure}


\end{document}